\documentclass[aps,prd,floatfix,nofootinbib,showpacs,twocolumn,10pt]{revtex4-1}

\usepackage{amssymb}
\usepackage[intlimits]{amsmath}
\usepackage{amsfonts}
\usepackage{dsfont}
\usepackage{subfigure}
\usepackage[usenames,dvipsnames]{color}

\definecolor{webgreen}{rgb}{0,0.75,0}
\definecolor{webred}{rgb}{0.75,0,0}
\definecolor{webblue}{rgb}{0,0,0.75}
\definecolor{darkblue}{rgb}{0,0,0.6}
\definecolor{dunkelgrau}{rgb}{0.8,0.8,0.8}
\definecolor{lgray}{rgb}{0.95,0.95,0.95}
\definecolor{lgreen}{rgb}{0.95,1.00,0.90}
\definecolor{lblue}{rgb}{0.9,0.95,1.00}
\definecolor{lred}{rgb}{1.00,0.90,0.80}
\definecolor{shadecolor}{rgb}{1.00,0.92,0.82}

\usepackage[colorlinks=true,linkcolor=darkblue,citecolor=darkblue,urlcolor=darkblue]{hyperref}
\usepackage{graphicx}

\usepackage{natbib}
\usepackage{multirow}
\usepackage{bbm}

        \usepackage[vcentermath]{youngtab}

      \def\longlongrightarrow{
      \relbar\joinrel\relbar\joinrel\relbar\joinrel\relbar\joinrel\rightarrow}

\begin{document}

\title{The three-gluon vertex in Landau gauge }
\author{Gernot Eichmann}
\email{gernot.eichmann@theo.physik.uni-giessen.de}
\affiliation{Institut f\"ur Theoretische Physik,  Justus-Liebig--Universit\"at Giessen, 35392 Giessen, Germany.}
\author{Richard Williams}
\email{richard.williams@theo.physik.uni-giessen.de}
\affiliation{Institut f\"ur Theoretische Physik, Justus-Liebig--Universit\"at Giessen, 35392 Giessen, Germany.}
\author{Reinhard Alkofer}
\email{reinhard.alkofer@uni-graz.at}
\affiliation{Institut f\"ur Physik, Karl-Franzens--Universit\"at Graz, Universit\"atsplatz 5, 8010 Graz, Austria.}
\author{Milan Vujinovic}
\email{milan.vujinovic@uni-graz.at}
\affiliation{Institut f\"ur Physik, Karl-Franzens--Universit\"at Graz, Universit\"atsplatz 5, 8010 Graz, Austria.}

\begin{abstract}
We present the first Dyson--Schwinger calculation of the three-gluon vertex in Landau-gauge QCD in which its full covariant structure is back-coupled self-consistently. We truncate a Bose-symmetrized version of the DSE at the level of one-loop diagrams, model the four-gluon vertex, and neglect terms that contain non-primitively divergent $n$-point functions; the ghost-gluon vertex is taken bare to good approximation. Fit functions for the ghost and gluon propagators that interpolate between scaling and decoupling are presented. In all aspects of our study Bose symmetry is manifest, from the truncation to the basis decomposition and to the momentum invariants.
We explore the uniform and soft-collinear infrared  limits and obtain the expected infrared exponents. The presence of a zero crossing in the tree-level component of the vertex is confirmed for both scaling- and decoupling-type scenarios.
The zero crossing appears at a scale $\sim 1$ GeV; however, its location might be sensitive to the four-gluon vertex and missing components in the DSE.

\end{abstract}

\maketitle

\section{Introduction}

In studies of Green's functions using the Dyson--Schwinger equations (DSEs)~\cite{Roberts:1994dr,Alkofer:2000wg,Fischer:2006ub},
the majority of effort has been placed on gluon and quark propagators
and how they couple via the quark-gluon vertex.
This necessitates knowledge of the other primitively divergent Green's functions,
such as the ghost propagator, ghost-gluon vertex, three-gluon vertex  and four-gluon vertex,
in addition to an infinite tower of higher $n$-point functions.

For propagators, there have been extensive studies within the DSE
and functional renormalization group approaches~\cite{vonSmekal:1997vx,Fischer:2002hna,Pawlowski:2003hq,Aguilar:2008xm,Fischer:2009tn},
together with direct comparisons with calculations on the lattice~\cite{Maas:2006qw,Sternbeck:2007ug,Cucchieri:2009zt,Sternbeck:2013zja,Cucchieri:2013nja}. Typically, though DSE calculations have employed Ans\"atze for the vertices, they have been fairly successful qualitatively which is suggestive that quantitative agreement is within reach. That we are close enables the use of fit functions to be employed in place of extended systems of coupled integral equations, thus minimising one technical complication in the process of `moving up the tower'.

Since we are interested primarily in hadronic properties derived from QCD, the connection between the gauge and matter fields is of paramount interest. Whilst  this coupling is explicitly mitigated through the quark-gluon vertex, it itself satisfies a DSE that induces an implicit dependence upon other $n$-point functions. Of particular interest are the three- and four-gluon vertices since they typify the self-interacting non-Abelian character of Yang-Mills theory and, when quarks are considered, QCD.

To date, the ghost-gluon vertex, quark-gluon vertex, three-gluon vertex and four-gluon vertex have been tackled (to some extent) in Landau gauge both functionally and on the lattice~\cite{Skullerud:2002ge,Alkofer:2008tt,Alkofer:2008dt,Cucchieri:2008qm,Kellermann:2008iw,Huber:2012kd,Rojas:2013tza,Ahmadiniaz:2013rla}. However, bar the ghost-gluon vertex, no full self-consistent DSE calculation in which the full covariant structure of the considered vertex is back-coupled has been completed.

The three-gluon vertex is an important input for phenomenological applications.
It has been explored in the context of gauge-invariance~\cite{Kim:1979ep,Ball:1980ax,Binosi:2011wi} and perturbation theory~\cite{Davydychev:2001uj,Binger:2006sj} and more recently it has been the focus of intense study~\cite{Ahmadiniaz:2013rla,Aguilar:2013vaa,Pelaez:2013cpa,Blum:2014gna}.  Lattice calculations of the three-gluon vertex in two and three dimensions give clear evidence that the leading tree-level component features a zero crossing at some infrared (IR) momentum scale~\cite{Maas:2007uv,Cucchieri:2008qm}. Though the $4$-dimensional studies are inconclusive, they are at least suggestive of a similar feature. Its presence, and in particular its location, may have profound effects upon a wealth of hadronic observables.
In particular, it has applications in meson spectroscopy beyond rainbow-ladder~\cite{Fischer:2009jm},
excited states, gluonic components of exotic mesons, hybrids and glueballs.
It also provides the irreducible three-body force in baryons which has so far not been considered
beyond the Faddeev equation with two-quark interactions~\cite{Eichmann:2009qa}
or its simplification to quark-diquark models~\cite{Oettel:1998bk,Oettel:2000jj,Eichmann:2008ef}.
Therein lie important questions such as two- vs. three-quark dominance in excited states
and the nature of baryonic hybrids~\cite{Klempt:2009pi,Dudek:2012ag}.
The three-gluon vertex is further relevant for the near-conformal window of QCD and QCD-like theories,
thus far only explored for propagators of strongly coupled theories \cite{August:2013jia}; it enters the quark-gluon vertex that is expected to drive the theory from a confining to a conformal phase.

In this paper, we study the structural properties of the three-gluon vertex
through a permutation group analysis following from Bose symmetry. We thus
establish the importance of the tensor components beyond tree-level.
In the sub-leading components we are able to resolve singularities that occur
when the momentum of one gluon becomes soft. These complement the usual divergence
in the uniform limit, whose power-law (logarithmic) nature depends upon the scaling (decoupling)
of the ghost propagator. We confirm the presence of the zero-crossing in the leading component
of the three-gluon vertex seen in similar studies and lattice calculations.
We demonstrate that a self-consistent DSE solution can shift its location from the deep IR toward a `hadronic' scale of $\sim 1$ GeV.
That value will depend on the truncation, thus indicating that the impact of the four-gluon vertex and missing diagrams should be explored in detail.
We also calculate the non-perturbative running coupling associated with the three-gluon vertex and determine its IR fixed point in the case of scaling.

The paper is organised as follows. In Sec.~\ref{sec:dse} we outline the DSE for the three-gluon vertex,
its Bose symmetrization and truncation, together with the ghost and gluon propagators and four-gluon vertex used as input.
In Sec.~\ref{sec:bose} we discuss Bose symmetry in detail, and the constraints it imposes on the tensor decomposition of the three-gluon vertex and the symmetry properties of the phase space.
In Sec.~\ref{sec:results} we present our results, including a summary of our numerical methods.
Finally we conclude in Sec.~\ref{sec:conclusions}. Further details regarding tensor bases are relegated to appendices.

\section{Three-gluon vertex DSE} \label{sec:dse}

\begin{figure}[t]
                  \begin{center}
                  \includegraphics[scale=0.12]{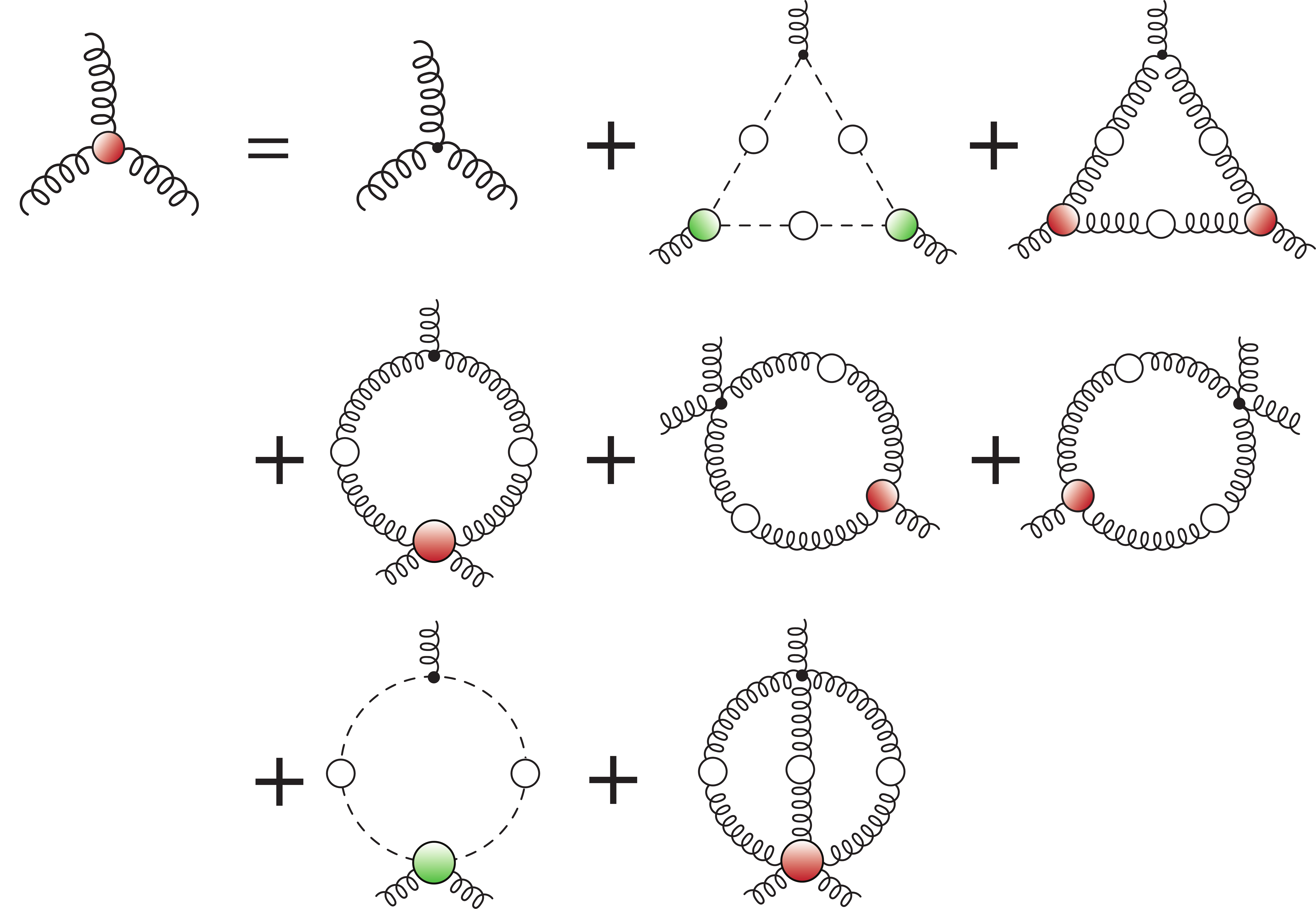}
                  \caption{The full DSE for the three-gluon vertex in QCD without quarks.
                           All dressed vertices are 1PI, except for the gluon five-point function in the last row
                           which contains further skeleton graphs including the 1PI three-gluon, four-gluon and five-gluon vertices.
                           }\label{fig:dse3g}
                  \end{center}
      \end{figure}

\begin{figure*}[t]
                  \begin{center}
                  \includegraphics[scale=0.17]{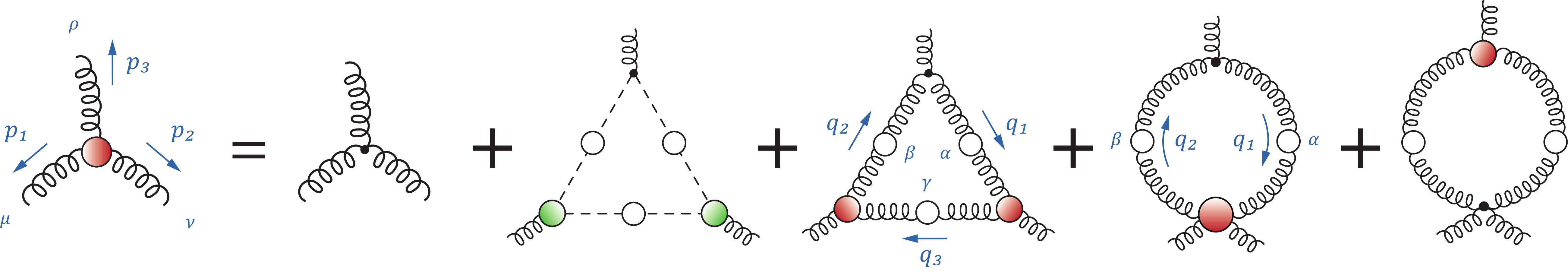}
                  \caption{The truncated DSE for the three-gluon vertex, together with the momentum routing used in Eqs.~(\ref{dse:ghostloop}--\ref{dse:sf2}). After symmetrization,
                           the resulting equation is identical to the symmetrized first two lines of Fig.~\ref{fig:dse3g}.}\label{fig:dse3gtrunc}
                  \end{center}
      \end{figure*}

       The full DSE for the three-gluon vertex in the standard one-particle irreducible (1PI) formulation is shown in Fig.~\ref{fig:dse3g}.
       It contains:
       \begin{itemize}
       \item the ghost and gluon loops from the first row;
       \item the `swordfish' diagrams in the second row, where the first depends on the
             dressed four-gluon vertex and the remaining two on the dressed three-gluon vertex;
       \item another ghost loop in the third row that depends on the ghost-gluon four-point function;
       \item and further two-loop terms which we absorbed in the last diagram. The gluon five-point function that appears here
             is a shorthand for skeleton graphs that contain the 1PI three-, four- and five-gluon vertices, see e.g. Ref.~\cite{Alkofer:2008dt}.
       \end{itemize}
       Incorporating quarks would produce two further diagrams analogous to the ghost loops.
      In the following we consider a truncation that neglects all two-loop diagrams
      and vertices without a tree-level counterpart, which leaves the top  two rows of Fig.~\ref{fig:dse3g}.
      In order to ensure Bose symmetry of the three-gluon vertex, we symmetrize the equation (which
      is equivalent to deriving the DSEs with respect to all three gluon legs and adding them together).
      The symmetrized sum of Fig.~\ref{fig:dse3g} is then identical to the symmetrized version of Fig.~\ref{fig:dse3gtrunc} which
      contains the ghost loop, the gluon loop, and two swordfish diagrams:
      \begin{equation}\label{3g-DSE}
      \Gamma^{\mu\nu\rho}_{3g} = \Gamma^{\mu\nu\rho}_{3g,0} + g^2 \left[ \Lambda^{\mu\nu\rho}_\text{(gh)}  + \Lambda^{\mu\nu\rho}_\text{(gl)}
                                                                       + \Lambda^{\mu\nu\rho}_\text{(sf,1)}  + \Lambda^{\mu\nu\rho}_\text{(sf,2)} \right].
      \end{equation}
      The diagrams are worked out explicitly in Table~\ref{tab:DSE}.

      The DSE depends on the ghost and gluon propagators, the ghost-gluon vertex and the four-gluon vertex as an input.
        The ghost and gluon propagators in Landau gauge are given by
        \begin{equation}\label{ghost-gluon}
        D_G(p) = -\frac{G(p^2)}{p^2} \,, \quad
        D^{\mu\nu}(p)  = \frac{Z(p^2)}{p^2}\,T^{\mu\nu}_p ,
        \end{equation}
        where $G(p^2)$ and $Z(p^2)$ are the scalar ghost and gluon dressing functions,
        $T_p^{\mu\nu}=\delta^{\mu\nu}-p^\mu p^\nu/p^2$ is the transverse \mbox{projector} with respect to the momentum $p$,
        and we will occasionally refer to $D(p^2)=Z(p^2)/p^2$ as the gluon `propagator'.

      The three-gluon vertex will always be fully contracted with gluon propagators when it appears, for example, in hadronic matrix elements.
      The transversality of the gluon in Landau gauge entails that only the transverse projection of the vertex can contribute to physical observables,
      and therefore it is sufficient to restrict oneself to the transverse projection of the vertex DSE.
      In practice, the DSE decouples into a transverse equation and longitudinal ones. The latter contain the transverse vertex solution as an input,
      but not vice versa, and thereby decouple from the dynamics~\cite{Fischer:2008uz}.
      All ingredients of Eq.~\eqref{3g-DSE} are therefore understood
      to be transversely projected. The ghost-gluon and four-gluon vertices require an explicit transverse projection, cf.~Table~\ref{tab:tree-level}.
      For the three-gluon vertex we only need to take into account the subset of transverse
      tensor structures which are discussed in Sec.~\ref{sec:bose-symm-tensor-basis} and App.~\ref{sec:orthonormal-tensor-basis}.

         \renewcommand{\arraystretch}{1.2}

             \begin{table*}[t]

                \normalsize

                \begin{align}
                    \Lambda_\text{(gh)} ^{\mu\nu\rho}(p_1,p_2,p_3) &= -N_C \int_q D_G(q_1^2) \, D_G(q_2^2) \, D_G(q_3^2)\,
                                                                       \Gamma^{\rho}_{gh,0}(-q_2,q_1,p_3) \, \Gamma^{\nu}_{gh}(-q_1,q_3,p_2) \, \Gamma^{\mu}_{gh}(-q_3,q_2,p_1)\,,  \label{dse:ghostloop}\\[2mm]
                    \Lambda^{\mu\nu\rho}_\text{(gl)}(p_1,p_2,p_3) &= \frac{N_C}{2} \int_q D(q_1^2) \, D(q_2^2) \, D(q_3^2)\,
                                                                      \Gamma^{\beta\alpha\rho}_{3g,0}(-q_2,q_1,p_3)\,\Gamma^{\alpha\gamma\nu}_{3g}(-q_1,q_3,p_2)\,\Gamma^{\gamma\beta\mu}_{3g}(-q_3,q_2,p_1)\,,\\[2mm]
                    \Lambda_\text{(sf,1)} ^{\mu\nu\rho}(p_1,p_2,p_3) &=  -\frac{3N_C}{4}\,\int_q D(q_1^2) \, D(q_2^2) \,
                                                                         \Gamma^{\beta\alpha\rho}_{3g,0}(-q_2,q_1,p_3)\,\Gamma^{\mu\nu\beta\alpha}_{4g}(p_1,p_2,q_2,-q_1)\,,  \\[2mm]
                    \Lambda_\text{(sf,2)} ^{\mu\nu\rho}(p_1,p_2,p_3) &=  -\frac{3N_C}{2} \int_q D(q_1^2) \, D(q_2^2) \,   \Gamma^{\beta\alpha\rho}_{3g}(-q_2,q_1,p_3)\,\Gamma^{\mu\nu\beta\alpha}_{4g,0}(p_1,p_2,q_2,-q_1)\,. \label{dse:sf2}
                \end{align}

               \caption{Diagrams in the three-gluon vertex DSE of Fig.~\ref{fig:dse3gtrunc}. The coefficients combine all color factors, tree-level prefactors from the QCD Lagrangian,
                        and multiplicities that arise in the DSE derivation. To be specific: ghost and gluon loops get a color factor $-\tfrac{N_C}{2}$ and swordfish diagrams $\tfrac{3N_C}{2}$;
                        the ghost loop has a symmetry factor $2$ and the swordfish diagrams $\tfrac{1}{2}$. The second swordfish picks up another factor 2 because it is counted twice in the symmetrization (cf.~Fig.~\ref{fig:dse3g}). The tree-level prefactors $\Gamma_\text{gh} \sim -ig$, $\Gamma_\text{3g} \sim ig$ and $\Gamma_\text{4g} \sim -g^2$ are factored out in the end, so that the tree-level vertices
                        take the form given below. All vertices are transverse in the gluon legs, and $\int_q = d^4q/(2\pi)^4$. }
               \label{tab:DSE}

                \begin{align}
                    \Gamma_\text{gh,0}^{\rho}(p_1,p_2,p_3) &= \widetilde{Z}_1\,T^{\rho\gamma}_{p_3}\left[\frac{p_2-p_1}{2}\right]^\gamma\,, \\[2mm]
                    \Gamma_\text{3g,0}^{\mu\nu\rho}(p_1,p_2,p_3) &= Z_1 \,T^{\mu\alpha}_{p_1}\,T^{\nu\beta}_{p_2}\,T^{\rho\gamma}_{p_3}\left[ \,\delta^{\alpha\beta}\,(p_1-p_2)^\gamma + \delta^{\beta\gamma}\,(p_2-p_3)^\alpha + \delta^{\gamma\alpha}\,(p_3-p_1)^\beta \,\right]\,,  \label{tree-level-3g}\\[2mm]
                    \Gamma_\text{4g,0}^{\mu\nu\rho\sigma}(p_1,p_2,p_3,p_4) &= Z_4\,T^{\mu\alpha}_{p_1}\,T^{\nu\beta}_{p_2}\,T^{\rho\gamma}_{p_3}\,T^{\sigma\delta}_{p_4}\left[ \,\delta^{\alpha\gamma}\,\delta^{\beta\delta} - \delta^{\alpha\delta}\,\delta^{\beta\gamma} \,\right] \,.  \label{tree-level-4g}
                \end{align}

               \caption{Transversely projected tree-level vertices that appear in the three-gluon vertex DSE.
                        The tree-level four-gluon vertex is the effective Lorentz structure that remains in the swordfish diagrams
                        after working out the color traces.}
               \label{tab:tree-level}

        \end{table*}

      In Ref.~\cite{Blum:2014gna} the coupled Yang-Mills system of ghost, gluon and three-gluon vertex DSEs was solved
      by retaining the dominant tensor structure of the three-gluon vertex.
      The goal of our study is complementary: we retain the full structure of the vertex and explore the impact of the remaining tensor components.
      In exchange, we use a fixed propagator input and treat the three-gluon vertex DSE as a standalone equation.
      This is justified from the analysis of Ref.~\cite{Blum:2014gna} where the back-reaction of the three-gluon vertex upon the propagator level was found to be small.
      The ghost and gluon propagators are reasonably well known in Landau gauge, hence we construct parametrizations for them which are detailed in Sec.~\ref{sec:propagators}.
      We replace the dressed ghost-gluon vertex by its tree-level form; it resembles the vertex DSE solution reasonably well~\cite{Cucchieri:2004sq,Schleifenbaum:2004id,Cucchieri:2006tf,Cucchieri:2008qm,Huber:2012kd}.
      For the four-gluon vertex we employ a tree-level model that reproduces the correct IR exponent and ultraviolet (UV) behavior, see Sec.~\ref{sec:four-gluon-vertex}.

\subsection{Renormalization}

    \renewcommand{\arraystretch}{1.5}

      We briefly discuss the renormalization constants that appear in Table~\ref{tab:tree-level}.
      The Yang-Mills sector of QCD contains five primitively divergent Green functions and hence five renormalization constants, plus one for the coupling $g$,
      that relate the renormalized with the bare quantities:
      \begin{equation}\label{GFs-dressed-vs-bare}
      \begin{array}{rl}
          G &= G^{(0)}/\widetilde{Z}_3\,, \\
          Z &= Z^{(0)}/Z_3\,, \\
          g &= g^{(0)}/Z_g\,,
      \end{array}\quad
      \begin{array}{rl}
           \Gamma_\text{gh}  &=  \Gamma_\text{gh}^{(0)} \widetilde{Z}_1 \,, \\
           \Gamma_\text{3g}  &=  \Gamma_\text{3g}^{(0)} Z_1\,, \\
           \Gamma_\text{4g}  &=  \Gamma_\text{4g}^{(0)} Z_4\,.
      \end{array}
      \end{equation}
      The Slavnov-Taylor identities relate the vertex renormalization constants
      to those of the propagators and the coupling via
      \begin{equation}\label{Z-relations-1}
         \widetilde{Z}_1 = Z_g\,Z_3^{1/2}\,\widetilde{Z}_3\,, \quad
         Z_1 = Z_g\,Z_3^{3/2}\,, \quad
         Z_4 = Z_g^2\,Z_3^2\,.
      \end{equation}
      Taylor's non-renormalization argument~\cite{Taylor:1971ff} states that the ghost-gluon vertex can stay unrenormalized in Landau gauge.
      Hence we can set $\widetilde{Z}_1=1$, which defines the MiniMOM scheme~\cite{vonSmekal:2009ae,Sternbeck:2012qs}, and all renormalization constants can be expressed through $\widetilde{Z}_3$ and~$Z_3$ :
      \begin{equation}\label{Z-relations-2}
         Z_g = \frac{1}{Z_3^{1/2} \widetilde{Z}_3}\,, \quad
         Z_1 = \frac{Z_3}{\widetilde{Z}_3}\,, \quad
         Z_4 = \frac{Z_3}{\widetilde{Z}_3^2}\,.
      \end{equation}
      As a consequence, all DSEs in the Yang-Mills sector are already renormalized
      once $\widetilde{Z}_3$ and $Z_3$ are known, so we do not need to set another renormalization condition for the three-gluon vertex.
      In practice $\widetilde{Z}_3$ and $Z_3$ are determined in the process of solving the ghost and gluon DSEs, cf. Sec.~\ref{sec:propagators} for a brief discussion.

      It is a simple check to confirm that the renormalization constants
      for the various diagrams (\ref{dse:ghostloop}--\ref{dse:sf2}) and~\eqref{tree-level-3g} in the three-gluon vertex DSE
      combine correctly. If we extract the intrinsic dependencies of all propagators and vertices on the renormalization constants
      according to Eq.~\eqref{GFs-dressed-vs-bare} and combine them in front of the integrals,
      \begin{itemize}
      \item the tree-level term provides a factor $Z_1$,
      \item the ghost loop gives $g^2/\widetilde{Z}_3^3 = g_0^2\,Z_3/\widetilde{Z_3} = g_0^2 \, Z_1$,
      \item the gluon loop: $g^2 \,Z_1^3 / Z_3^3= g_0^2 \,Z_1$,
      \item and the swordfish diagrams: $g^2 \,Z_1  Z_4/Z_3^2 = g_0^2 \,Z_1$,
      \end{itemize}
      and therefore $\Gamma_\text{3g} = \Gamma_\text{3g}^{(0)} Z_1$ holds.

         \renewcommand{\arraystretch}{1.2}

             \begin{table}[t]

                \begin{center}
                \begin{tabular}{  @{\;\;} l @{\;\;\;\;}     @{\;\;\;}r@{\;\;\;}   @{\;\;\;}r@{\;\;\;}   @{\;\;\;\;}r@{\;\;}      }

                           \hline\noalign{\smallskip}

                                         & SC         &  DC   & UV        \\  \noalign{\smallskip}\hline\noalign{\smallskip}

                    $\Gamma_\text{gh}$   & $0$           &  $0$     & $0$     \\
                    $G^{-1}$                  & $\kappa$     &  $0$     & $-\delta=\tfrac{9}{44}$     \\
                    $Z^{-1}$                  & $-2\kappa$     &  $-1$     & $1+2\delta=\tfrac{13}{22}$     \\
                    $\Gamma_\text{3g}$   & $-3\kappa$    &  $0$     & $1+3\delta=\tfrac{17}{44}$     \\
                    $\Gamma_\text{4g}$   & $-4\kappa$    &  $0$     & $1+4\delta=\tfrac{2}{11}$     \\

                           \noalign{\smallskip}\hline

                \end{tabular}
                \end{center}

               \caption{IR and UV exponents of the primitively divergent Green functions in Yang-Mills theory.
                        The IR power is the exponent of $p^2$ (modulo potential logarithms) after removing the canonical dimension. `SC' denotes scaling and `DC' decoupling.
                        In the scaling case, the vertices can have further soft-gluon singularities in the IR~\cite{Alkofer:2008jy,Alkofer:2008dt,Fischer:2009tn}.
                        The UV anomalous dimension is the exponent of $\ln p^2$.}
               \label{tab:GFs-IR-UV}

        \end{table}

    \renewcommand{\arraystretch}{1.2}

\subsection{Ghost and gluon propagators}  \label{sec:propagators}

        \renewcommand{\arraystretch}{1.2}

        Here we provide details on our parametrizations for the ghost and gluon propagators.
        The data sets correspond to the calculation in Ref.~\cite{Fischer:2002hna}, where the Yang-Mills system
        was solved upon neglecting two-loop terms in the gluon DSE and using tree-level ghost-gluon and three-gluon vertices; the latter was augmented by a dressing.
        In addition to the scaling solution discussed in that work, we also use four sets of decoupling solutions
        obtained in the same truncation.\footnote{We are grateful to C.~S.~Fischer for providing us with these data.}
        Parametrizations for ghost and gluon propagators are available from the literature, but they were either designed
        to fit the scaling solution of the DSEs~\cite{Fischer:2003rp} or decoupling solutions obtained on the lattice~\cite{Aguilar:2010cn}.
        In order to study both scenarios, we will construct parametrizations below
        that can interpolate between these cases.
        Furthermore, we also wish to implement features that were recently obtained
        via a direct DSE solution of the Yang-Mills system in the complex plane~\cite{Strauss:2012dg}.
        In that study the only non-analytic structure of the ghost dressing function was found to be a cut on the time-like axis,
        whereas the gluon exhibited an additional peak at $p^2 = -\Lambda^2$, with $\Lambda\sim 0.6 \dots 0.7$~GeV.

        For simplicity we restrict ourselves to parametrizations where the IR, mid-momentum and UV parts factorize,
        since this simplifies the separate discussion of the IR and UV behavior.
        We work with a single gluonic mass scale $\Lambda=0.6$ GeV
        and express all subsequent formulas through the dimensionless variable $x=p^2/\Lambda^2$.
        The resulting parametrizations have the form
        \begin{equation}\label{GZ-fits}
        \begin{split}
            G(x) &= G_\text{IR}(x) \,G_\text{M}(x)\,G_\text{UV}(x)\,, \\[1mm]
            Z(x) &= Z_\text{IR}(x) \,Z_\text{M}(x)\,Z_\text{UV}(x)\,,
        \end{split}
        \end{equation}
        where $G_\text{IR}$, $Z_\text{IR}\rightarrow 1$ in the UV and $G_\text{UV}$, $Z_\text{UV}\rightarrow 1$ in the IR.

        For the IR behavior we employ functions which interpolate between the scaling and decoupling type.
        In the scaling scenario, the ghost dressing diverges with $x^{-\kappa}$ and the gluon dressing vanishes with $x^{2\kappa}$. In the decoupling
        case, the ghost dressing $G(0)$ is constant and that of the gluon vanishes with $x$, so that the gluon propagator $D(0)$ becomes constant in the IR.
        The simplest way to accommodate both cases, without altering the UV or introducing time-like poles,
        is to work with powers of the function
        \begin{equation}
            s(x,a) = \frac{x^\kappa+a}{x^\kappa+1}\,,
        \end{equation}
        where the parameter $a$ discriminates between scaling ($a=0$) and decoupling ($a>0$).
        The IR parts of ghost and gluon dressing functions are then constructed as
        \begin{equation}\label{propagator-fits-IR}
        \begin{split}
            G_\text{IR}(x) &=  s(x,a_1)^{-1}\,, \\
            Z_\text{IR}(x) &= \frac{x}{x+1}\,s(x,a_2)^{2-\frac{1}{\kappa}} \,,
        \end{split}
        \end{equation}
        where the scaling exponent is $\kappa\simeq0.595$~\cite{Lerche:2002ep,Zwanziger:2001kw} and the fit parameters $a_1$, $a_2$ are given in Table~\ref{tab:GZ-fits}.

        For the mid-momentum and UV behavior we employ the function
        \begin{equation}\label{h-function}
        \begin{split}
            h(x,c) &= \frac{1}{c x + \ln x} - \frac{x_0}{\left(1-\ln x_0\right)(x-x_0)}\,, \\
            x_0 &= e^{-W(c)} = \frac{W(c)}{c}\,,
        \end{split}
        \end{equation}
        where $W(c)$ is the product logarithm or Lambert-$W$ function, the solution of the equation $W(c)\,e^{W(c)}=c$.
        Since the (inverse) zero of $c x+\ln x$ at $x=x_0$ has been expanded around $x_0$ and subtracted in the second term,
        $h(x,c)$ is analytic except for a branch cut extending from $x=0$ to minus infinity. In the IR $h(x,c)$ goes to a constant,
        \begin{equation}
            h(x,c) \stackrel{x\rightarrow 0}{\longlongrightarrow} \frac{1}{1-\ln x_0}\,,
        \end{equation}
        whereas for large $x$ it is suppressed, either with an inverse power of $x$ (if $c>0$) or logarithmically ($c=0$).
        The latter case is useful for modelling the UV running of the ghost and gluon dressing functions without altering their IR behavior, since
        for $c=0$ Eq.~\eqref{h-function} reduces to
        \begin{equation}\label{better-logs}
            h(x,0) = \frac{1}{\ln x} - \frac{1}{x-1}  \;\; \stackrel{x\rightarrow 0}{\longlongrightarrow}  \;\; 1\,,
        \end{equation}
        which follows from $W(c\rightarrow 0) = c + \dots \Rightarrow x_0=1$.
        Hence, for
        the UV parts in Eq.~\eqref{GZ-fits} we use
        \begin{equation}
        \begin{split}
            G_\text{UV}(x) &= b_1\,h(x,0)^\frac{9}{44}\,, \\
            Z_\text{UV}(x) &= b_2\,h(x,0)^\frac{13}{22}\,,
        \end{split}
        \end{equation}
        and determine the parameters $b_1=1.10$ and $b_2=1.32$
        from the UV running of the dressing functions at large~$x$:
        \begin{equation}
            G(x)  \rightarrow \frac{b_1}{(\ln x)^\frac{9}{44}}\,, \quad
            Z(x) \rightarrow \frac{b_2}{(\ln x)^\frac{13}{22}}\,.
        \end{equation}
        For the mid-momentum parts $G_M(x)$ and $Z_M(x)$ we obtained reasonable fits with the following functional forms:
        \begin{equation}\label{GZ-fits-2}
        \begin{split}
            G_\text{M}(x) &= 1 + \frac{c_1 + d_1 \left[ x^\kappa h(x,\tfrac{1}{2})\right]^2 }{1+x^\kappa}\,, \\
            Z_\text{M}(x) &= 1 + \frac{c_2 \left[x^\kappa h(x,\tfrac{1}{2}) \right] + d_2 \left[ x^\kappa h(x,\tfrac{1}{2})\right]^2 }{1+x^\kappa}\,,
        \end{split}
        \end{equation}
        where $c=\tfrac{1}{2}$ in Eq.~\eqref{h-function} leads to $x_0\simeq0.703467$.
        The remaining parameters are
        \begin{equation}\label{fit-parameters}
            \begin{array}{rl}
               c_1 &= 0.81 + 1.42 \, a_1\,, \\
               c_2 &= 2.45 - 5.12 \, a_2\,,
            \end{array}\quad
            \begin{array}{rl}
               d_1 &= -6.85\,, \\
               d_2 &= 28.5\,.
            \end{array}
        \end{equation}

         \renewcommand{\arraystretch}{1.2}

             \begin{table}[t]

                \begin{center}
                \begin{tabular}{  @{\;\;} c @{\;\;\;\;}     @{\;\;\;}c@{\;\;}   @{\;\;}c@{\;\;\;}  @{\;\;\;}c@{\;\;}   @{\;\;}c@{\;\;\;}   @{\;\;}c@{\;\;}     }

                           \hline\noalign{\smallskip}

                           Set     & $a_1$    &  $a_2$   & $G(0)$ & $D(0)\,\Lambda^2$  & $\widetilde{Z}_3$       \\

                           \noalign{\smallskip}\hline\noalign{\smallskip}

                    1 (SC)        & $0$       &  $0$     & $\infty$  & $0$            & $1.529$     \\
                    2 (DC)        & $0.02$    &  $0.03$  & $100$     & $0.41$         & $1.519$     \\
                    3 (DC)        & $0.24$    &  $0.26$  & $10$      & $0.86$        & $1.429$     \\
                    4 (DC)        & $0.58$    &  $0.42$  & $5$       & $1.00$        & $1.329$     \\
                    5 (DC)        & $1.38$    &  $0.41$   & $3$       & $0.99$       & $1.195$      \\

                           \noalign{\smallskip}\hline

                \end{tabular}
                \end{center}

               \caption{Fit parameters $a_1$ and $a_2$ for the five data sets (`SC' = scaling, `DC' = decoupling).
               The ghost dressing and (dimensionless) gluon propagator at vanishing momentum as inferred from the data sets are collected as well.
               The last column shows the ghost renormalization constant $\widetilde{Z}_3$ for the various sets.}
               \label{tab:GZ-fits}

        \end{table}

    \renewcommand{\arraystretch}{1.2}

       \begin{figure*}[t]
                  \begin{center}
                  \includegraphics[scale=0.26]{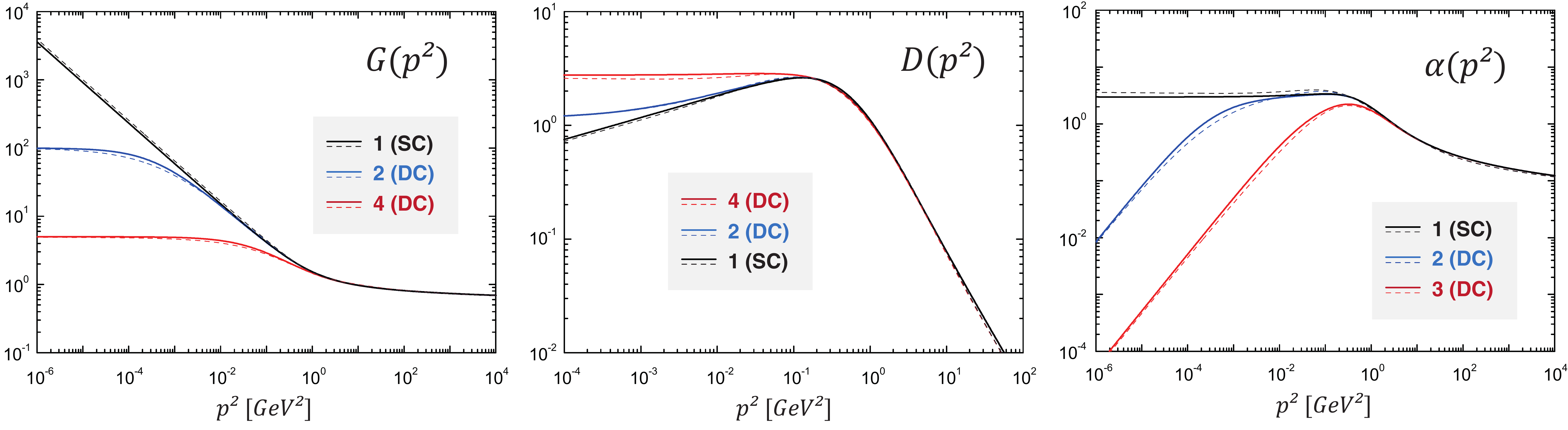}
                  \caption{Ghost dressing, gluon propagator, and
                           renormalization-point independent running coupling $\alpha(p^2) = \alpha(\mu^2) Z(p^2) G^2(p^2)$.
                           The DSE results (solid curves) are compared to our fits (dashed curves) for the scaling case and two decoupling sets.
                           The legends follow the plots from top to bottom.}\label{fig:propagator-fits}
                  \end{center}
      \end{figure*}

        We did not attempt to include the dependence on the renormalization parameters in our fits.
        The renormalization procedure is described in detail in Refs.~\cite{Lerche:2002ep,Fischer:2002hna}.
         In order to solve the ghost and gluon DSEs, one has to specify the coupling $\alpha(\mu^2)=g^2/(4\pi)$ and two
         boundary values for $Z(p^2)$ and $G(p^2)$. The former provides the connection with experiment and simultaneously sets the scale in GeV units;
         otherwise the only scale would be the numerical cutoff $\Lambda_\text{cutoff}$ that enters the equations.
         The boundary values for $Z(\mu^2)$ and $G(\mu^2)$ are (within certain constraints) arbitrary; their specification leads to subtracted, finite equations
         at the subtraction point $\mu$ which also determine $Z_3$ and $\widetilde{Z}_3$ in the process.
        The data sets correspond to a fixed choice $\mu = 2.28$ GeV, $\alpha(\mu^2)=0.7427$ and $\Lambda_\text{cutoff} = 316$ GeV.
        The resulting values for the ghost renormalization constant $\widetilde{Z}_3$ are listed in Table~\ref{tab:GZ-fits}, whereas
        the gluon renormalization constant $Z_3 = 3.384$ is the same for scaling and decoupling.

         While the choice of $Z(\mu^2)\,G^2(\mu^2)$ reflects the freedom of renormalization, the value of $G(\mu^2)$ discriminates between scaling and decoupling behavior in the IR.
         Since the ghost DSE determines the inverse ghost dressing function that approaches the perturbative limit from below,
         lowering $G(\mu^2)^{-1}$ leads to a limit where $G(0)^{-1}$ vanishes, which is the scaling solution.
         It is then numerically more convenient to subtract the ghost equation directly at $p^2=0$ and use $G(0)$ as the second boundary condition.
        The resulting values of $G(0)$ and $D(0)$ are also collected in Table~\ref{tab:GZ-fits}.
         In practice one additionally has to ensure the absence of spurious quadratic divergences and longitudinal artefacts in the gluon DSE
         which can arise due to the truncation; however, these issues are independent of the existence of scaling and decoupling solutions~\cite{Fischer:2008uz,Fischer:2010is}.

        The fits are shown in Fig.~\ref{fig:propagator-fits} for the scaling case and two decoupling solutions.
        They describe the data reasonably well over the whole momentum domain.
        We should note that we aimed for simplicity rather than precision: one could improve the quality of the fits by relaxing the linear dependence of
        the parameters $c_i$ on $a_i$ in Eq.~\eqref{fit-parameters}, or by altering the form of $G_\text{M}(x)$ and $Z_\text{M}(x)$, etc.
        We note that also the resulting spectral functions from the fits are in qualitative agreement with
        the direct DSE solutions from the complex-plane calculation in Ref.~\cite{Strauss:2012dg}.
        We included the pole $x/(x+1)$ in Eq.~\eqref{propagator-fits-IR} on purpose to obtain a peak in the gluon spectral function;
        one could replace this factor for example with $x h(x,c)$ to obtain parametrizations with time-like branch cuts only.

        Lowering the ghost dressing away from the scaling limit $G(0)\rightarrow\infty$
        leads to a nonzero, increasing gluon propagator $D(0)$.
        Our propagator fits yield:
        \begin{equation}
        \begin{split}
           G(0) &= \frac{b_1}{a_1}\,(1+c_1) = b_1\left(1.42+\frac{1.81}{a_1}\right)\,, \\
           D(0) &= \frac{Z(x)}{x\,\Lambda^2}\Big|_{x=0} =(a_2)^{2-\frac{1}{\kappa}}\,\frac{b_2}{\Lambda^2}\,.
        \end{split}
        \end{equation}
        It is interesting that they produce not only a maximum but also a minimum value for the ghost dressing function at zero momentum:
        if $a_1\rightarrow \infty$, then $G(0)$ goes to a constant $\approx 1.56$. At this point $D(0)$ has reached a plateau
        and slightly decreased again. If one writes $D(0) = 1/m^2$ and interprets $m$ as an effective `gluon mass',
        then in this `extreme' decoupling case one has from Table~\ref{tab:GZ-fits}: $m \approx \Lambda$, whereas in the scaling limit $m\rightarrow\infty$.

\subsection{Four-gluon vertex} \label{sec:four-gluon-vertex}

       \begin{figure}[!b]
                  \begin{center}
                  \includegraphics[scale=0.31]{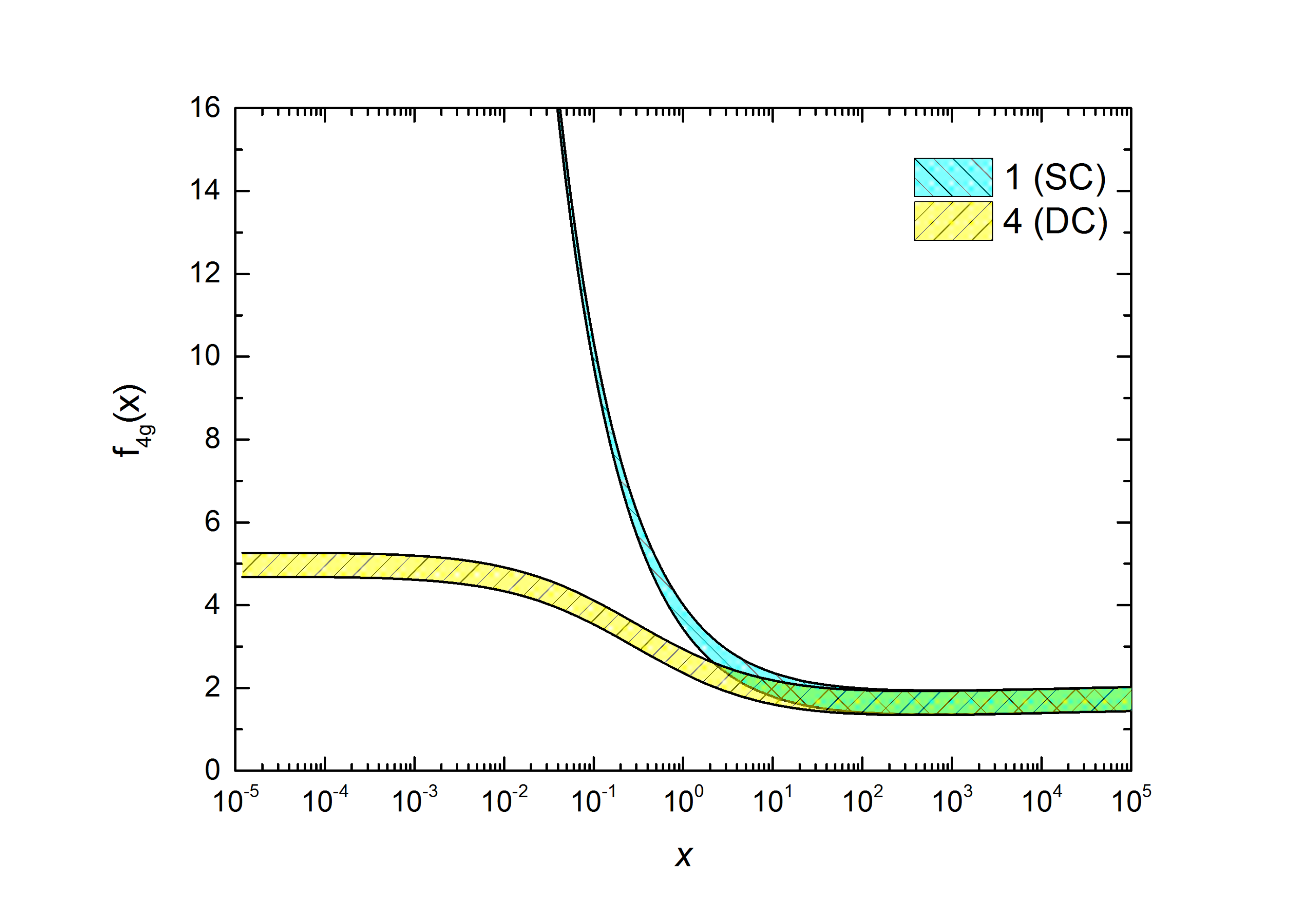}
                  \caption{Four-gluon vertex dressing of Eq.~\eqref{4g-dressing}.}\label{fig:4gv-dressing}
                  \end{center}
      \end{figure}

      Whereas the Landau-gauge ghost and gluon propagators and the ghost-gluon vertex have been studied extensively in the past,
      the non-perturbative properties of the dressed four-gluon vertex~\cite{Baker:1976vz} are still largely \textit{terra incognita}.  The one-loop perturbative behavior is known~\cite{Celmaster:1979km,Brandt:1985zz,Papavassiliou:1992ia}, and non-perturbative explorations have been made~\cite{Stingl:1994nk,Driesen:1997wz,Driesen:1998xc}.
      In Ref.~\cite{Kellermann:2008iw} the four-gluon vertex DSE in the scaling case was solved by retaining the three Lorentz-color tensor structures
      which are momentum independent and Bose-symmetric.
      From Table~\ref{tab:GFs-IR-UV}, the UV anomalous dimension of the four-gluon vertex is $1+4\delta=\tfrac{2}{11}$.
      In the IR the vertex goes to a constant (decoupling) or scales with $-4\kappa$.
      We implement these features here by a simple Ansatz:
      \begin{equation}
         \Gamma_\text{4g}^{\mu\nu\rho\sigma}(p_1,p_2,p_3,p_4) = f_\text{4g}(x)\,\Gamma_\text{4g,0}^{\mu\nu\rho\sigma}\,,
      \end{equation}
      where $\Gamma_\text{4g,0}^{\mu\nu\rho\sigma}$ is the tree-level tensor structure.
      Upon transverse projection and implementation in the swordfish diagrams, it takes the form of Eq.~\eqref{tree-level-4g}.
      The Bose-symmetric dressing function is modelled by
      \begin{equation}\label{4g-dressing}
         f_\text{4g}(x) = b_4\,\frac{h(x)^{-\frac{2}{11}}}{s(y,a_4)^4}\,, \quad x = \frac{p_1^2+p_2^2+p_3^2+p_4^2}{4\,\Lambda^2}\,,
      \end{equation}
      with $y=x/4$. In the scaling case we have again $a_4=0$; the choice $a_4=0.43 \,a_1$ yields a sensible extension to the decoupling case.
      The renormalization constant $Z_4$ is already implicit in the tree-level structure, and $f_\text{4g}(x)\rightarrow 1$
      at the numerical cut-off entails $b_4=0.63$.
      In order to study the dependence of our results on the four-gluon vertex model, we shift it by a constant: $f_\text{4g}(x) + (0 \dots 0.6)$.
      This produces the band shown in Fig.~\ref{fig:4gv-dressing}.

\section{Bose symmetry} \label{sec:bose}

        \renewcommand{\arraystretch}{0.8}

         The permutation group is a powerful tool for exploring the structure properties of the three-gluon vertex.
         The full vertex including momentum, Lorentz and color parts is Bose-symmetric.
         Since the color structure $f_{abc}$ is totally antisymmetric,
         the combination of Lorentz and momentum parts must be antisymmetric as well:
             \begin{equation}\label{symmetry-relations-3g-vertex}
             \begin{split}
                   \Gamma^{\mu\nu\rho}(p_1,p_2,p_3) &= -\Gamma^{\nu\mu\rho}(p_2,p_1,p_3) \\
                 = \Gamma^{\nu\rho\mu}(p_2,p_3,p_1) &= -\Gamma^{\rho\nu\mu}(p_3,p_2,p_1) \\
                 = \Gamma^{\rho\mu\nu}(p_3,p_1,p_2) &= -\Gamma^{\mu\rho\nu}(p_1,p_3,p_2)\,.
             \end{split}
             \end{equation}
         It can be decomposed in 14 tensor structures with Lorentz-invariant dressing functions:
        \begin{equation*}
           \Gamma^{\mu\nu\rho}(p_1,p_2,p_3) = \sum_{i=1}^{14}f_i(p_1^2,p_2^2,p_3^2)\,\tau_i^{\mu\nu\rho}(p_1,p_2,p_3)\,.
        \end{equation*}
         The Bose symmetry property allows one to arrange both the tensor basis of the vertex and its dressing functions
         into irreducible multiplets of the permutation group $\mathds{S}^3$, and subsequently combine them to obtain antisymmetric product representations.
         As we will see below, from the permutation group analysis one can already make a number of statements about the symmetry properties of the phase space,
         and hence the expected momentum dependence of the vertex dressing functions.

\subsection{Kinematics}

        \renewcommand{\arraystretch}{1.3}

\begin{figure}[t]
                    \begin{center}
                    \includegraphics[scale=0.25]{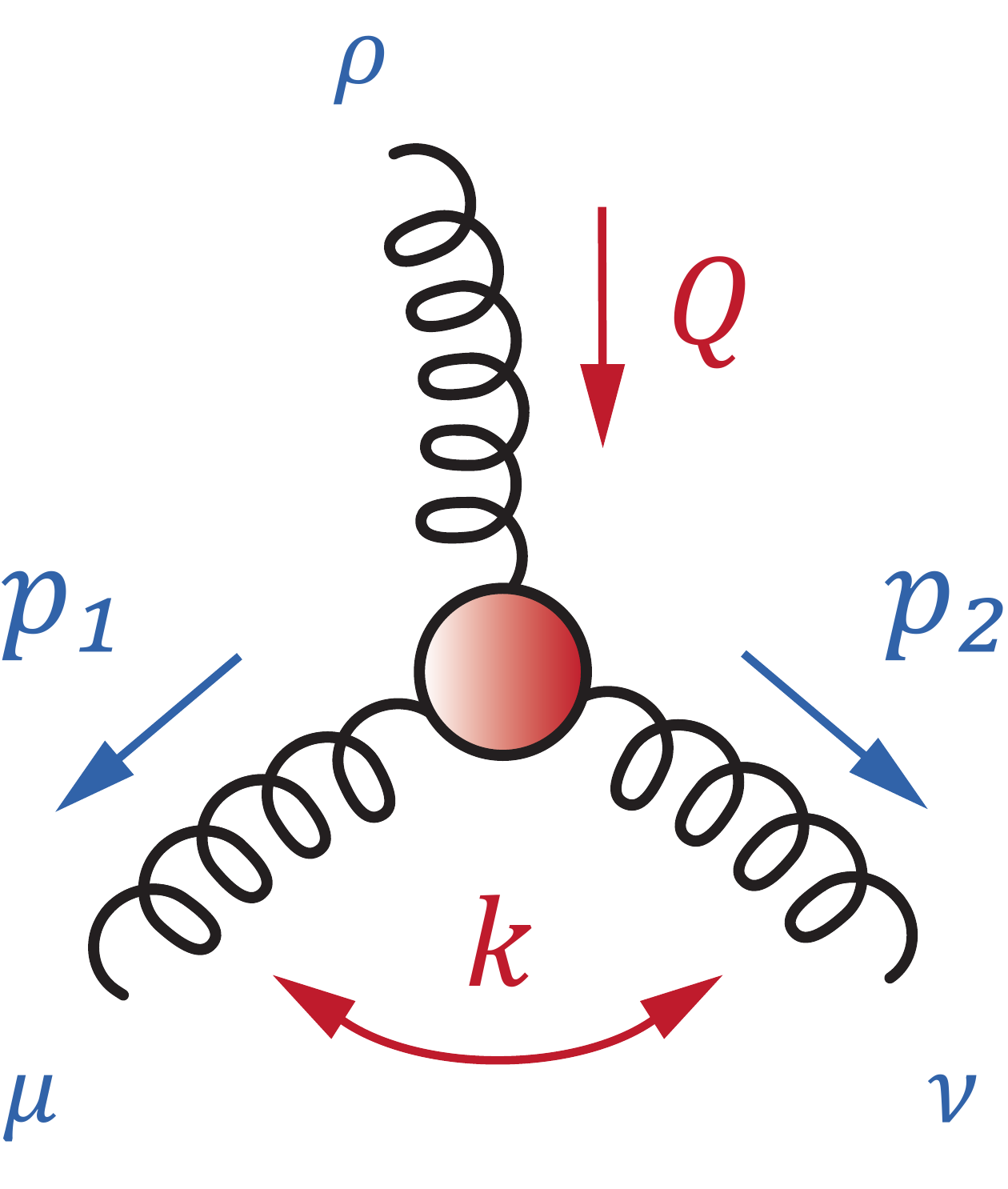}
                    \caption{ Momentum routing in the three-gluon vertex.
                                            }\label{fig:3gvertex}
                    \end{center}
        \end{figure}

             Since only two of the momenta in the three-gluon vertex are independent,
             it is useful to work with the combinations  (cf. Fig.~\ref{fig:3gvertex})
             \begin{equation}\label{momenta-relative-total}
                   k = \frac{p_2-p_1}{2}\,, \qquad
                   Q = -p_3\,,
             \end{equation}
             instead of $p_1$, $p_2$ and $p_3$, so that
             \begin{equation}
                   p_1 = -k+\frac{Q}{2}\,, \quad
                   p_2 = k+\frac{Q}{2}\,, \quad
                   p_3 = -Q.
             \end{equation}
             If we write $\Gamma^{\mu\nu\rho}(p_1,p_2,p_3) = \Gamma^{\mu\nu\rho}(k,Q)$ and define
             \begin{equation}
                 \begin{array}{rl}
                     k' &= -\tfrac{1}{2}\,(k+\tfrac{3Q}{2}), \\
                     k'' &= -\tfrac{1}{2}\,(k-\tfrac{3Q}{2}),
                 \end{array}\qquad
                 \begin{array}{rl}
                     Q' &= k - \tfrac{Q}{2}\,,\\
                     Q'' &= -k-\tfrac{Q}{2}\,,
                 \end{array}
             \end{equation}
             the symmetry relations in Eq.~\eqref{symmetry-relations-3g-vertex} take the form
             \begin{alignat}{2}
                  & \,\Gamma^{\mu\nu\rho}(k,Q) &&= -\Gamma^{\nu\mu\rho}(-k,Q) \nonumber \\
                 =& \,\Gamma^{\nu\rho\mu}(k',Q') &&= -\Gamma^{\rho\nu\mu}(-k',Q') \\
                 =& \,\Gamma^{\rho\mu\nu}(k'',Q'') &&= -\Gamma^{\mu\rho\nu}(-k'',Q'')\,. \nonumber
             \end{alignat}

             From $k$ and $Q$ one can construct three Lorentz-invariants $k^2$, $Q^2$ and $k\cdot Q$. We express them for convenience in terms of the variables
             \begin{equation}\label{three-LI}
                 t = \frac{Q^2}{4}\,, \qquad \xi = \frac{4k^2}{3Q^2}\,, \qquad z = \widehat{k}\cdot\widehat{Q}\,,
             \end{equation}
             where only $t$ carries a dimension. The hats denote normalized four-momenta. In the space-like DSE calculation,
             $t$ and $\xi$ are real and positive whereas $z\in[-1,1]$ is the cosine of the polar angle.
             Below we will form combinations of $t$, $\xi$ and $z$ which are multiplets of the permutation group.

        \renewcommand{\arraystretch}{0.8}

\subsection{Permutation-group multiplets}

         We generically denote multiplets that transform
        under irreducible representations of the permutation group by\footnote{Some of the following
        discussion and notation is based on Refs.~\cite{Carimalo:1992ia,Eichmann:2011vu}, where
        the nucleon's Faddeev amplitude was analysed in an analogous fashion.}
        \begin{equation}\label{S3-multiplets-generic}
            \mathcal{S}\,, \quad
            \mathcal{A}\,, \quad
            \mathcal{D}_1 = \left[ \begin{array}{c} a_1 \\ s_1 \end{array}\right],\quad
            \mathcal{D}_2 = \left[ \begin{array}{c} a_2 \\ s_2 \end{array}\right],
        \end{equation}
        or, in terms of Young tableaux:
        \begin{equation*}
            {\yng(3)\;, \qquad \yng(1,1,1)\;, \qquad  \yng(2,1)\;, \qquad  \yng(2,1)} \;.
        \end{equation*}
         $\mathcal{S}$ and $\mathcal{A}$ are completely symmetric or antisymmetric singlets, and each doublet $\mathcal{D}_i$
        has a mixed-antisymmetric entry $a_i$
        and a mixed-symmetric component $s_i$.

        To make this explicit, consider a function of three momenta $\psi(p_1,p_2,p_3)$.
        $\psi$ is completely generic and can represent a Lorentz-invariant momentum variable or dressing function,
        a four-momentum,
        or also a given tensor basis element. We first define the combinations
        \begin{equation}\label{psi-1-3}
        \begin{split}
            \psi_1^\pm &:= \psi(p_1,p_2,p_3) \pm \psi(p_2,p_1,p_3)\,, \\
            \psi_2^\pm &:= \psi(p_2,p_3,p_1) \pm \psi(p_1,p_3,p_2)\,,   \\
            \psi_3^\pm &:= \psi(p_3,p_1,p_2) \pm \psi(p_3,p_2,p_1)\,,
        \end{split}
        \end{equation}
        which are mixed-symmetric or -antisymmetric under exchange of the momentum indices $1$ and $2$.
        The singlets and doublets generated from $\psi(p_1,p_2,p_3)$
        are then the following linear combinations:
        \renewcommand{\arraystretch}{1.2}
        \begin{equation}\label{perm-multiplets}
        \begin{split}
            \mathcal{S} &= \psi_1^+ + \psi_2^+ + \psi_3^+ \,, \\
            \mathcal{A} &= \psi_1^-  + \psi_2^- + \psi_3^- \,, \\[2mm]
            \mathcal{D}_1
             &= \left[ \begin{array}{c}
                                    \psi_2^--\psi_3^- \\
                                    -\tfrac{1}{\sqrt{3}} \left(  \psi_2^+ + \psi_3^+ -2\psi_1^+ \right) \end{array}\right] ,\\
            \mathcal{D}_2
            & = \left[ \begin{array}{c}
                                    \tfrac{1}{\sqrt{3}} \left( \psi_2^- + \psi_3^- -2\psi_1^- \right) \\
                                    \psi_2^+-\psi_3^+ \end{array}\right] .
        \end{split}
        \end{equation}
        $\mathcal{S}$ and $\mathcal{A}$ are fully symmetric or antisymmetric under exchange of momenta.
        The doublet entries $a_i$, $s_i$ are (anti-) symmetric with respect to the \textit{indices} 1 and 2.
        The whole doublet $\mathcal{D}_1$ is symmetric under exchange of the first two momentum \textit{arguments}
        whereas $\mathcal{D}_2$ is antisymmetric.

        The doublets transform under the two-dimensional (orthogonal) matrix representations of $\mathds{S}^3$ which are given by
        \begin{equation}\label{S3-rep-matrices} \renewcommand{\arraystretch}{0.8}
           \mathsf{M} = \left(\begin{array}{rr} -1 & \, 0 \\ 0 & \ 1 \end{array}\right), \qquad
           \mathsf{M}_\pm = \frac{1}{2}\left(\begin{array}{cc} 1 & \pm \sqrt{3} \\ \pm \sqrt{3} & -1 \end{array}\right)\,.
        \end{equation}
        This can be verified from the six permutation operators
        \begin{equation}\label{permutation-operators}
            1\,, \quad
            P_{12}\,, \quad
            P_{13}\,, \quad
            P_{23}\,, \quad
            P_{23}\,P_{12}\,, \quad
            P_{13}\,P_{12}\,,
        \end{equation}
        which are understood to act on the indices of the arguments $p_i$ (instead of interchanging their positions),
        for example: $P_{23}\,P_{12}\,\psi(p_1,p_2,p_3) = \psi(p_3,p_1,p_2)$.
        The three combinations in Eq.~\eqref{psi-1-3} follow if one applies
        \begin{equation}
           1 \pm P_{12}\,, \quad P_{31}\,P_{12} \pm P_{23}\,, \quad P_{32}\,P_{21} \pm P_{13}\,,
        \end{equation}
        to $\psi(p_1,p_2,p_3)$.
        Using the relations $P_{ij}=P_{ji}$, $P_{ij}^2=1$, and $P_{ij}\,P_{jk}=P_{jk}\,P_{ki}=P_{ki}\,P_{ij}$ (without summation),
        one can show that the doublets $\mathcal{D}_i$ transform as \renewcommand{\arraystretch}{1.2}
        \begin{equation}\label{permutation-tf}
        \begin{array}{rl}
           P_{12}\, \mathcal{D}_i &= \mathsf{M}\,\mathcal{D}_i \,, \\
           P_{13}\,\mathcal{D}_i  &= \mathsf{M}_+\,\mathcal{D}_i\,, \\
           P_{23}\,\mathcal{D}_i  &= \mathsf{M}_-\,\mathcal{D}_i\,,
        \end{array} \qquad
        \begin{array}{rl}
           P_{13}\,P_{12}\,\mathcal{D}_i  &= \mathsf{M}\,\mathsf{M}_+\,\mathcal{D}_i\,, \\
           P_{23}\,P_{12}\,\mathcal{D}_i  &= \mathsf{M}\,\mathsf{M}_-\,\mathcal{D}_i\,.
        \end{array}
        \end{equation}
        For a given permutation operator,
        both doublets $\mathcal{D}_1$, $\mathcal{D}_2$ transform under its same irreducible representation;
        hence they form a two-dimensional irreducible subspace.

        \renewcommand{\arraystretch}{0.8}

        Our notation makes it particularly simple to study product representations, which we will need in the following.
        In order to obtain a symmetric singlet in the product space, one can either combine two doublets or two (symmetric or antisymmetric) singlets:
        \begin{equation}\label{perm-1}
           \mathcal{D}\cdot\mathcal{D}' :=aa'+ss' \,, \qquad  \mathcal{S}\mathcal{S}'\,, \qquad  \mathcal{A}\mathcal{A}'\,.
        \end{equation}
        The singlet property of $\mathcal{D}\cdot\mathcal{D}'$ follows from the orthogonality of the representation matrices in Eq.~\eqref{S3-rep-matrices}.
        Similarly, antisymmetric singlets are constructed from
        \begin{equation}\label{perm-2}
           \mathcal{D} \times \mathcal{D'} := as'-sa' \,, \qquad  \mathcal{S}\mathcal{A}\,.
        \end{equation}
        Doublets are obtained from the trivial combination $\mathcal{S}\mathcal{D}$, but also from
        \begin{equation}\label{perm-3}
            \mathcal{D} \ast \mathcal{D}':= \left[ \begin{array}{c} as'+sa' \\ aa'-ss' \end{array}\right] , \quad
            \mathcal{D} \ast \mathcal{A}  := \left[ \begin{array}{c} s \\ -a \end{array}\right] \mathcal{A}\,.
        \end{equation}
        One can show for example that $\mathcal{D} \ast \mathcal{D}'$ satisfies the same transformation properties as in Eq.~\eqref{permutation-tf}.
        We collect some useful identities along the way:
        \begin{equation}
        \begin{split}
            \mathcal{D}\ast(\mathcal{D}\ast\mathcal{D}') &= (\mathcal{D}\cdot\mathcal{D})\,\mathcal{D}'\,, \\
            \mathcal{D}\times(\mathcal{D}\ast\mathcal{D}') &= -(\mathcal{D}\ast\mathcal{D})\times\mathcal{D}'\,,\\
            \mathcal{D}\times(\mathcal{D}\ast\mathcal{A}) &= -(\mathcal{D}\cdot\mathcal{D})\ast\mathcal{A}'\,.
        \end{split}
        \end{equation}
        The relation
        \begin{equation}\label{perm-decomposition-doublet}
            \bigg[\frac{(\mathcal{D}\ast\mathcal{D})\times\mathcal{D}'}{(\mathcal{D}\ast\mathcal{D})\times\mathcal{D}}\bigg]\,\mathcal{D}
            - \bigg[\frac{\mathcal{D}\times\mathcal{D}'}{(\mathcal{D}\ast\mathcal{D})\times\mathcal{D}}\bigg]\,\mathcal{D}\ast\mathcal{D}
             = \mathcal{D}'
        \end{equation}
        states that a doublet $\mathcal{D}'$ can be expanded in two doublets $\mathcal{D}$ and $\mathcal{D}\ast\mathcal{D}$
        and thereby related to two totally symmetric singlets (the numerators and denominators in the brackets are totally antisymmetric).

        In the discussion below, $\mathcal{D}$ will usually operate in the space of Lorentz invariants and $\mathcal{D}'$ in the space of Lorentz tensors.
        For illustration, take Eq.~\eqref{perm-2}: if $a$ and $s$ are two Lorentz invariants
        which form a permutation-group doublet, and if $a'$ and $s'$ denote two tensor basis elements which also form a doublet,
        then the combination $as'-as'$ is a new tensor basis element that is fully antisymmetric.
        Since the total vertex (modulo color) must be antisymmetric as well, the corresponding dressing function can only
        depend on fully symmetric Lorentz invariants. We will construct such variables in the following.

        \renewcommand{\arraystretch}{1.2}

\subsection{Bose-symmetric Lorentz invariants} \label{sec:lorentz-invariants}

        To begin with, let us first arrange the momenta that enter the three-gluon vertex in the multiplet structure.
        We can use the four-momentum $\psi(p_1,p_2,p_3)=p_3$ as the permutation-group `seed' and write:
        \begin{equation}
        \begin{split}
            \psi(p_1,p_2,p_3) = \psi(p_2,p_1,p_3) = p_3\,, \\
            \psi(p_2,p_3,p_1) = \psi(p_3,p_2,p_1) = p_1\,,\\
            \psi(p_3,p_1,p_2) = \psi(p_1,p_3,p_2) = p_2 \,,
        \end{split}
        \end{equation}
        and Eq.~\eqref{psi-1-3} leads to
        \begin{equation}
            \begin{array}{rl}
                \psi_1^+ &= -2Q\,, \\
                \psi_2^+ &= Q\,, \\
                \psi_3^+ &= Q\,,
            \end{array}
            \qquad
            \begin{array}{rl}
                \psi_1^- &= 0\,, \\
                \psi_2^- &= -2k\,, \\
                \psi_3^- &= 2k.
            \end{array}
        \end{equation}
        The resulting $\mathcal{S}$, $\mathcal{A}$ and $\mathcal{D}_2$ from Eq.~\eqref{perm-multiplets} are all zero, and only one doublet remains:
        \begin{equation}\label{momenta-doublet}
            \mathcal{D}_1 \sim \left[\begin{array}{c} \tfrac{1}{\sqrt{3}}\,k \\ \tfrac{1}{2}\,Q \end{array}\right].
        \end{equation}
        These are the two independent momenta of the three-gluon vertex that make the permutation-group features most transparent.

        \renewcommand{\arraystretch}{1.3}

          \begin{figure}[t]
                    \begin{center}
                    \includegraphics[scale=0.18]{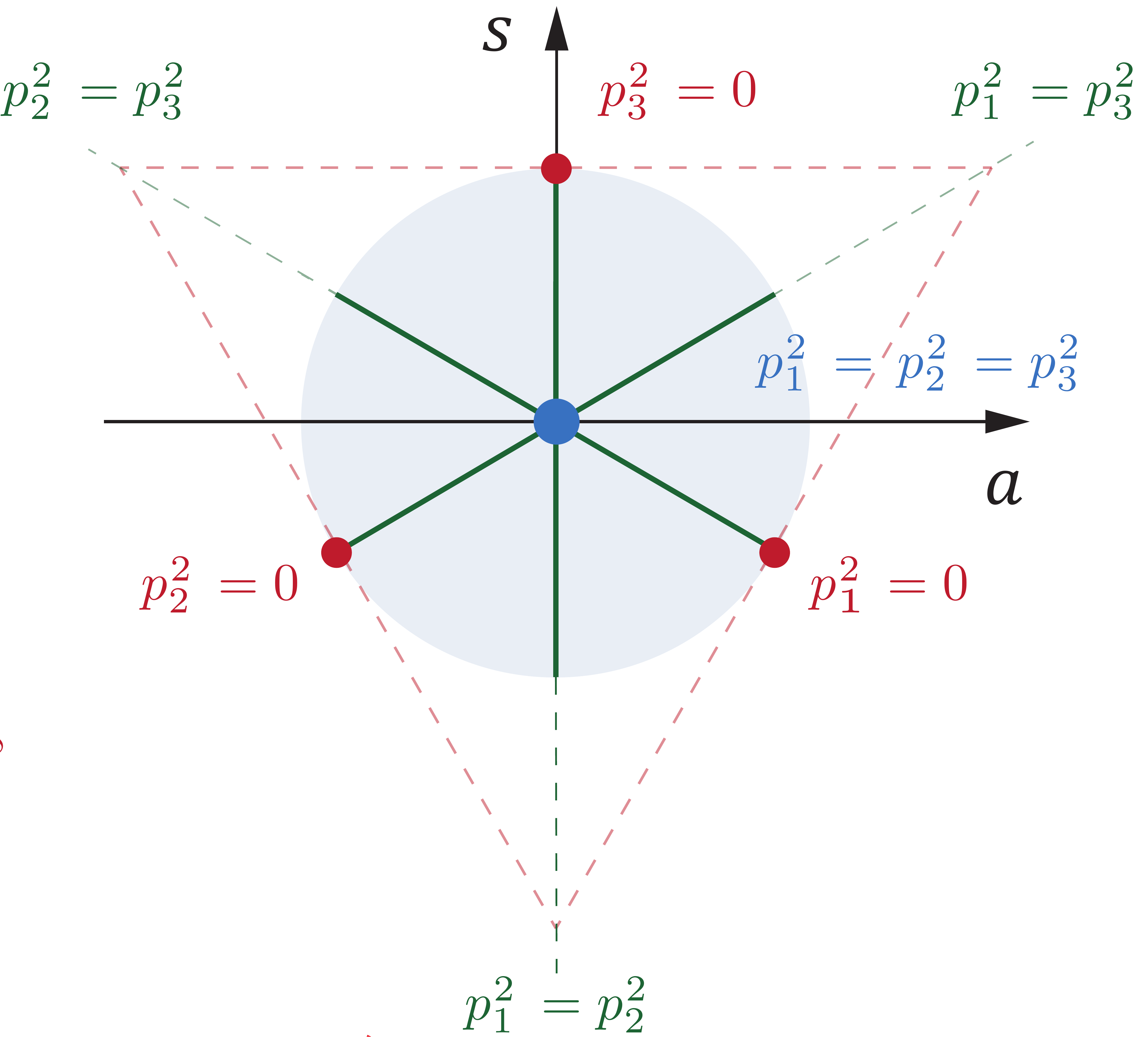}
                    \caption{ Phase space for the three-gluon vertex in the $(a,s)$ plane at a slice of fixed $\mathcal{S}$.
                                            }\label{fig:phasespace}
                    \end{center}
        \end{figure}

        $\mathcal{D}_1$ can be used to construct Lorentz-invariant variables in the product space.
        It follows from Eqs.~(\ref{perm-1}--\ref{perm-3}) that from one doublet one can only get a
        singlet ($\mathcal{D}_1 \cdot \mathcal{D}_1$) and another doublet ($\mathcal{D}_1 \ast \mathcal{D}_1$);
        the antisymmetric combination $\mathcal{D}_1\times\mathcal{D}_1$ vanishes. The resulting three Lorentz invariants,
        if we interpret the operations $\cdot$, $\times$ and $\ast$ as scalar products of four-momenta, are:
        \begin{equation}
        \begin{split}
            \mathcal{S}_0 &= \mathcal{D}_1 \cdot \mathcal{D}_1 = \frac{k^2}{3} + \frac{Q^2}{4} = t\left(1+\xi\right) \,, \\
            \mathcal{D} &= \frac{\mathcal{D}_1 \ast \mathcal{D}_1}{\mathcal{S}_0}
                         = \frac{1}{\mathcal{S}_0}\left[\begin{array}{c} \frac{k\cdot Q}{\sqrt{3}} \\ \frac{k^2}{3} - \frac{Q^2}{4} \end{array}\right]
                         = \left[\begin{array}{c} \frac{2\sqrt{\xi}\,z}{\xi+1} \\ \frac{\xi-1}{\xi+1}  \end{array}\right],
        \end{split}
        \end{equation}
        where $t$, $\xi$ and $z$ are the invariants defined in Eq.~\eqref{three-LI}.
        We divided $\mathcal{D}_1 \ast \mathcal{D}_1$ by the singlet $\mathcal{S}_0$ in order to remove its mass dimension.
        In this way we arrived at one dimensionful variable $\mathcal{S}_0 \in \mathds{R}_+$ and two dimensionless angular variables.
        To keep the notation simple, we will henceforth refer to them plainly as
        \begin{equation}\label{three-LI-S3}
            \mathcal{S}_0 = t\left(1+\xi\right)\,, \quad a =  \frac{2\sqrt{\xi}\,z}{\xi+1}\,, \quad s = \frac{\xi-1}{\xi+1}\,,
        \end{equation}
        with the inverse relations
        \begin{equation}
            t = \frac{1-s}{2}\,\mathcal{S}_0\,, \quad \xi = \frac{1+s}{1-s}\,, \quad z = \frac{a}{\sqrt{1-s^2}}\,.
        \end{equation}
        Expressed in terms of $p_1^2$, $p_2^2$ and $p_3^2$, they are given by
        \begin{equation}\label{S0-a-s}
        \begin{split}
             \mathcal{S}_0 = \frac{1}{6}\,&(p_1^2+p_2^2+p_3^2)\,, \\
            a = \sqrt{3}\,\frac{p_2^2-p_1^2}{p_1^2+p_2^2+p_3^2}&\,,  \quad
            s = \frac{p_1^2+p_2^2-2p_3^2}{p_1^2+p_2^2+p_3^2}\,,
        \end{split}
        \end{equation}
        which we could have also obtained directly
        by starting from $\psi(p_1,p_2,p_3)=p_3^2$ and repeating the steps~(\ref{psi-1-3}--\ref{perm-multiplets}).

        For fixed $\mathcal{S}_0>0$, the resulting phase space in the $(a,s)$ plane is the interior of a unit circle, illustrated in Fig.~\ref{fig:phasespace}.
        This follows from $z\in [-1,1]$ and \mbox{$\xi >0$ $\Rightarrow$ $a^2+s^2 \leq 1$} for $|z| \leq 1$. Adding the $\mathcal{S}_0$ direction,
        the space-like region which is sampled in the three-gluon vertex DSE becomes a cylindrical tube with unit radius.
        In Fig.~\ref{fig:phasespace} we show various momentum configurations in the $(a,s)$ plane; they are all independent of the symmetric variable $\mathcal{S}_0$:
        \begin{itemize}
        \item The symmetric limit $p_1^2=p_2^2=p_3^2$ is the origin of the $(a,s)$ plane.
              In terms of the variables~\eqref{three-LI}: $k^2=\tfrac{3}{4}\,Q^2$ and $k\cdot Q=0$ $\Rightarrow$ $\xi=1$ and $z=0$.
        \item The three soft kinematic limits where only one of the gluon momenta vanishes constitute a triangle: 
              \begin{equation}
              \begin{split}
                  p_1^2 = 0 \quad &\Leftrightarrow \quad s=-2+\sqrt{3}\,a \,, \\
                  p_2^2 = 0 \quad &\Leftrightarrow \quad s=-2-\sqrt{3}\,a \,, \\
                  p_3^2 = 0 \quad &\Leftrightarrow \quad s=1 \,.
              \end{split}
              \end{equation}
              It intersects with the unit circle at the three points \renewcommand{\arraystretch}{0.8}
              \begin{equation}
                  \left[\begin{array}{c} a \\s \end{array}\right] \; = \;
                       \frac{1}{2}\left[\begin{array}{c}  \sqrt{3} \\ -1 \end{array}\right], \;\;
                       -\frac{1}{2}\left[\begin{array}{c} \sqrt{3} \\ 1 \end{array}\right], \;\;
                       \left[\begin{array}{c} 0 \\1 \end{array}\right].
              \end{equation}
        \item The lines where two momenta coincide are also shown in the figure.
              They correspond to $a=\pm\sqrt{3}\,s$ or $a=0$ and intersect each other at the origin.
        \end{itemize}

        \renewcommand{\arraystretch}{1.3}

          \begin{figure}[t]
                    \begin{center}
                    \includegraphics[scale=0.18]{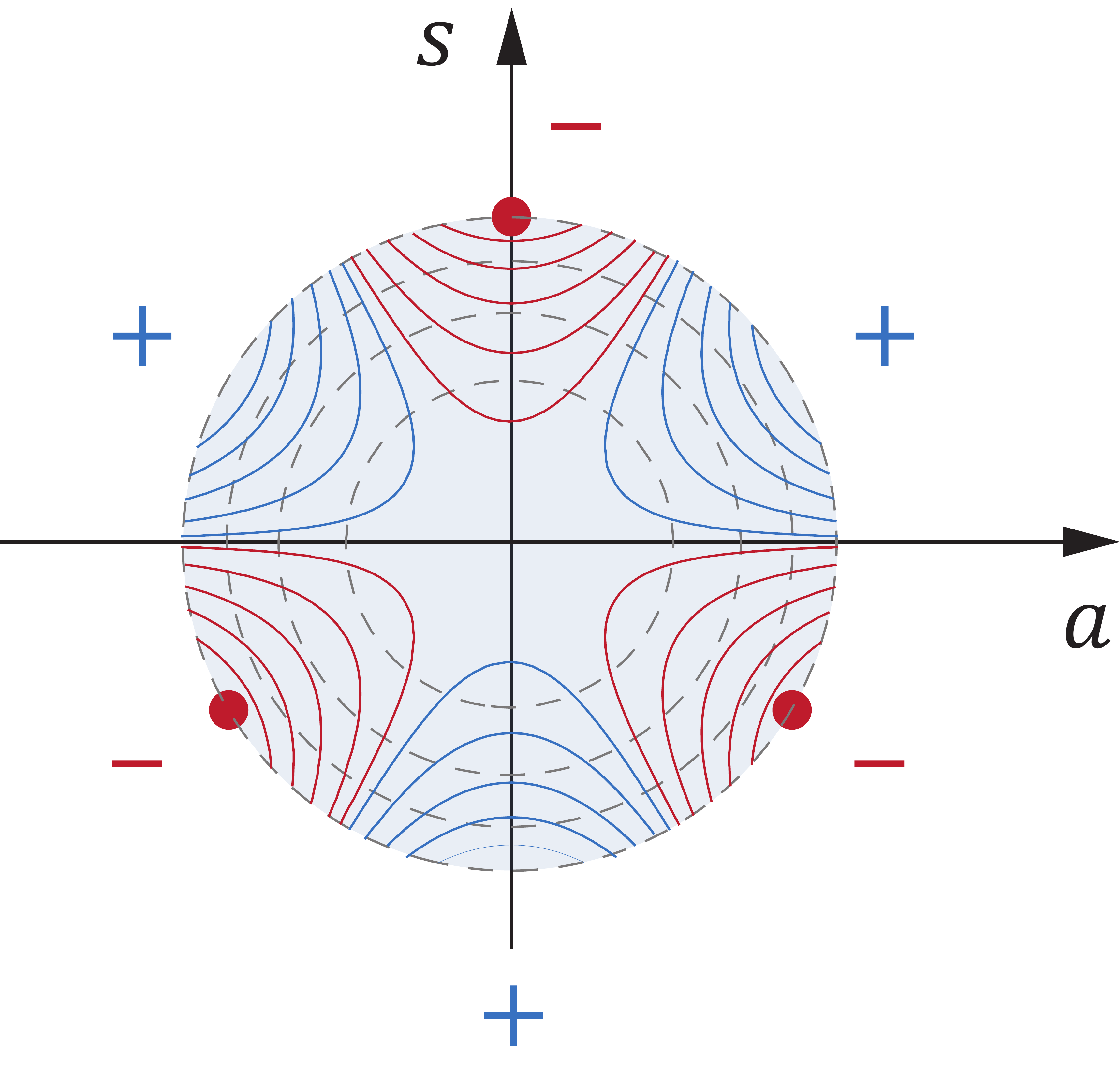}
                    \caption{ Contours of constant $\mathcal{S}_1$ (dashed circles) and constant $\mathcal{S}_2$ (solid curves) in the $(a,s)$ plane.
                              The signs indicate the regions where $\mathcal{S}_2$ is either positive or negative. At the three soft kinematic points, $\mathcal{S}_2=-1$.
                                            }\label{fig:phasespace2}
                    \end{center}
        \end{figure}

        Bose symmetry entails that each slice of $120^{\circ}$ in the $(a,s)$ plane carries the complete phase space information.
        To see this explicitly, suppose that we expand the vertex in a fully antisymmetric tensor basis. The corresponding dressing functions must be symmetric and can only
        depend on totally symmetric momentum variables.
        We can construct two such variables from $a$ and $s$, namely
        \begin{equation}\label{two-LI}
        \begin{split}
              \mathcal{S}_1 &=\mathcal{D}\cdot\mathcal{D} = a^2+s^2   \, \in \,[0,1]\,, \\
              \mathcal{S}_2 &=(\mathcal{D}\ast\mathcal{D})\cdot\mathcal{D} = 3a^2 s-s^3  \, \in \,[-1,1]\,.
        \end{split}
        \end{equation}
        All further combinations are redundant.
        The dressing functions of the vertex will then depend on $\mathcal{S}_0$, $\mathcal{S}_1$ and $\mathcal{S}_2$.
        The contours of constant $\mathcal{S}_1$ and $\mathcal{S}_2$ in the $(a,s)$ plane are shown in Fig.~\ref{fig:phasespace2}.
        $\mathcal{S}_1 \leq 1$ describes the radius of the circle, and the $\mathcal{S}_2$ profile exhibits the $120^{\circ}$ symmetry
        which must be reflected in the vertex dressing functions.
        All three soft kinematic limits correspond to the same value $\mathcal{S}_2=-1$.

        These observations are interesting in light of potential soft singularities of the three-gluon vertex.
        For the scaling solution, the vertex exhibits an IR singularity with $-3\kappa$ in the uniform momentum limit ($\mathcal{S}_0 = 0$)
        and further soft singularities in the kinematic limits $\mathcal{S}_2 = -1$~\cite{Alkofer:2008jy}.
        Since the scaling and decoupling solutions are continuously connected,
        we expect to encounter remnants of this singular behavior
        also in the decoupling solutions.
        In particular, we should recover the profiles of the dressing functions in Fig.~\ref{fig:phasespace2} in our numerical results,
        together with an enhancement at the three soft kinematic points.

        \renewcommand{\arraystretch}{1.4}

\subsection{Bose-symmetric tensor basis}  \label{sec:bose-symm-tensor-basis}

        In order to reveal the symmetry properties of the dressing functions, the corresponding
        tensor structures of the three-gluon vertex must be cast in permutation-group multiplets as well.
        The vertex has 14 Lorentz basis tensors.
        For convenience we work with the momenta
        \begin{equation}
            p_i \qquad \text{and} \qquad q_i = p_j-p_k\,,
        \end{equation}
        where the $p_i$ are the usual gluon momenta (which are outgoing in our convention) and $\{i,j,k\}$ is an even permutation of $\{1,2,3\}$.
        This seeming redundancy is resolved by the fact that only
        \begin{equation}
            p_1^\mu, \; p_2^\nu, \; p_3^\rho, \qquad  q_1^\mu, \; q_2^\nu, \; q_3^\rho\,,
        \end{equation}
        will appear in what follows, i.e., the momentum labels are intertwined with the respective Lorentz indices.

        The simplest construction principle starts from
        the following six seed elements which exhaust the possible momentum content in $p_i$ and $q_i$:
        \begin{equation}\label{psi123}
        \begin{array}{rl}
            \psi_1^{\mu\nu\rho} &= \tfrac{1}{2}\,\delta^{\mu\nu} \,q_3^\rho\,, \\
            \psi_2^{\mu\nu\rho} &= \tfrac{1}{6}\,q_1^\mu\, q_2^\nu\, q_3^\rho\,, \\
            \psi_3^{\mu\nu\rho} &= \tfrac{1}{2}\,q_1^\mu\, q_2^\nu\, p_3^\rho\,, \\
        \end{array}
        \quad
        \begin{array}{rl}
            \psi_4^{\mu\nu\rho} &= \tfrac{1}{2}\,p_1^\mu\, p_2^\nu\, q_3^\rho \,,\\
            \psi_5^{\mu\nu\rho} &= \tfrac{1}{6}\,p_1^\mu\, p_2^\nu\, p_3^\rho\,, \\
            \psi_6^{\mu\nu\rho} &= \tfrac{1}{2}\,\delta^{\mu\nu} \,p_3^\rho\,.
        \end{array}
        \end{equation}
        They are sufficient to generate a complete, linearly independent tensor basis.
        Following the steps~(\ref{psi-1-3}--\ref{perm-multiplets}),
        one can arrange their permutations into multiplets
        $\mathcal{A}'(\psi_i)$, $\mathcal{S}'(\psi_i)$, $\mathcal{D}_j'(\psi_i)$.
        The permutations act on the Lorentz indices and momentum labels of the $p_i$.
        We denote the multiplets
        with primes to distinguish them from the Lorentz invariants from the last section.
        The resulting 14 non-vanishing tensor elements are collected in Table~\ref{tab:permutation-group-basis}.
        $\mathcal{A}'(\psi_1)$ is the tree-level structure of the three-gluon vertex from Eq.~\eqref{tree-level-3g}.

         \renewcommand{\arraystretch}{1.2}

             \begin{table}[t]

                \begin{center}
                \begin{tabular}{  @{\;\;} c @{\qquad}      @{\quad}c@{\;\;}      }

                           \hline\noalign{\smallskip}

                    $\mathcal{A}'(\psi_1)$        & $q_1^\mu\,\delta^{\nu\rho} +q_2^\nu\,\delta^{\rho\mu} + q_3^\rho\,\delta^{\mu\nu}$          \\[1mm]
                    $\mathcal{D}_2'(\psi_1)$      & $\left[ \begin{array}{c} \frac{1}{\sqrt{3}}\left( q_1^\mu\,\delta^{\nu\rho} +q_2^\nu\,\delta^{\rho\mu} -2\, q_3^\rho\,\delta^{\mu\nu}\right)  \\
                                                                  q_1^\mu\,\delta^{\nu\rho} -q_2^\nu\,\delta^{\rho\mu} \end{array}\right]$      \\  \noalign{\medskip}\hline\noalign{\smallskip}

                    $\mathcal{A}'(\psi_2)$        & $q_1^\mu\,q_2^\nu\,q_3^\rho$      \\   \noalign{\medskip}\hline\noalign{\smallskip}

                    $\mathcal{S}'(\psi_3)$        & $p_1^\mu\,q_2^\nu\,q_3^\rho+q_1^\mu\,p_2^\nu\,q_3^\rho+q_1^\mu\,q_2^\nu\,p_3^\rho$    \\[1mm]

                    $\mathcal{D}_1'(\psi_3)$        & $\left[ \begin{array}{c} p_1^\mu\,q_2^\nu\,q_3^\rho-q_1^\mu\,p_2^\nu\,q_3^\rho  \\
                                                                  -\frac{1}{\sqrt{3}}\left( p_1^\mu\,q_2^\nu\,q_3^\rho+q_1^\mu\,p_2^\nu\,q_3^\rho -2\,q_1^\mu\,q_2^\nu\,p_3^\rho\right) \end{array}\right]$       \\ \noalign{\smallskip}\hline \noalign{\smallskip}

                    $\mathcal{A}'(\psi_4)$        & $q_1^\mu\,p_2^\nu\,p_3^\rho+p_1^\mu\,q_2^\nu\,p_3^\rho+p_1^\mu\,p_2^\nu\,q_3^\rho$    \\[1mm]
                    $\mathcal{D}_2'(\psi_4)$      & $\left[ \begin{array}{c} \frac{1}{\sqrt{3}}\left( q_1^\mu\,p_2^\nu\,p_3^\rho+p_1^\mu\,q_2^\nu\,p_3^\rho-2\,p_1^\mu\,p_2^\nu\,q_3^\rho\right)  \\
                                                                  q_1^\mu\,p_2^\nu\,p_3^\rho-p_1^\mu\,q_2^\nu\,p_3^\rho \end{array}\right]$        \\ \noalign{\smallskip}\hline \noalign{\smallskip}

                    $\mathcal{S}'(\psi_5)$        & $p_1^\mu\,p_2^\nu\,p_3^\rho$        \\  \noalign{\medskip}\hline\noalign{\smallskip}

                    $\mathcal{S}'(\psi_6)$        & $p_1^\mu\,\delta^{\nu\rho} +p_2^\nu\,\delta^{\rho\mu} + p_3^\rho\,\delta^{\mu\nu}$       \\[1mm]
                    $\mathcal{D}_1'(\psi_6)$        & $\left[ \begin{array}{c} p_1^\mu\,\delta^{\nu\rho} -p_2^\nu\,\delta^{\rho\mu}  \\
                                                                  -\frac{1}{\sqrt{3}}\left( p_1^\mu\,\delta^{\nu\rho} +p_2^\nu\,\delta^{\rho\mu} -2\, p_3^\rho\,\delta^{\mu\nu}\right) \end{array}\right]$     \\  \noalign{\medskip}\hline\noalign{\smallskip}

                \end{tabular}
                \end{center}

               \caption{Tensor basis for the three-gluon vertex in the permutation-group arrangement. The $p_i$ are the gluon momenta and $q_i=p_j-p_k$.}
               \label{tab:permutation-group-basis}

        \end{table}

        \renewcommand{\arraystretch}{0.8}

        The arrangement in Table~\ref{tab:permutation-group-basis} entails that all elements that contain
        either of the vectors $p_1^\mu$, $p_2^\nu$ or $p_3^\rho$ vanish upon a full transverse projection with $T_{p_1}^{\alpha\mu}\,T_{p_2}^{\beta\nu}\,T_{p_3}^{\gamma\rho}$.
        The three-gluon vertex in Landau gauge will always be contracted with such a transverse projector
        by virtue of the gluon propagators, and hence only the transverse part carries the dynamics.
        It is then sufficient to retain the four tensor structures that survive the projection:
        \begin{equation}
           \mathcal{A}'(\psi_1)\,, \quad \mathcal{A}'(\psi_2)\,, \quad \mathcal{D}_2'(\psi_1)\,.
        \end{equation}

        In principle one could exploit color gauge invariance and split each of the four remaining dressing functions into two parts:
        one which is fixed by the Slavnov-Taylor identity (STI) for the three-gluon vertex, and another one which is purely transverse and subject to analyticity constraints.
        The STI is not very helpful in practice because it depends on the unknown ghost-gluon scattering kernel.
        Nevertheless, contributions from both parts will generally survive the transverse projection.

        The basis elements $\mathcal{A}'(\psi_1)$ and $\mathcal{A}'(\psi_2)$ are already antisymmetric whereas those contained in the doublet $\mathcal{D}_2'(\psi_1)$ are not.
        We can combine the doublet with the Lorentz invariants $a$ and $s$ from Eq.~\eqref{three-LI-S3} to form further antisymmetric tensor structures.
        From Eqs.~\eqref{perm-2}, \eqref{perm-3} and \eqref{perm-decomposition-doublet} we have the following independent possibilities:
        \begin{equation}\label{D,D*D}
            \mathcal{D}\times\mathcal{D}_2'(\psi_1)\,, \qquad
            (\mathcal{D}\ast\mathcal{D})\times\mathcal{D}_2'(\psi_1)\,.
        \end{equation}
        In practice we find it useful to work with the linear combinations
        \begin{equation}\label{transverse-basis-final}
        \begin{split}
            \tau_1 &= \mathcal{A}'(\psi_1)\,, \\
            \tau_2 &= \frac{1}{\mathcal{S}_0}\,\mathcal{A}'(\psi_2)\,, \\
            \tau_3 &= 2\, \Big( \mathcal{A}'(\psi_1) -\tfrac{\sqrt{3}}{2}\,\mathcal{D}\times\mathcal{D}_2'(\psi_1)\Big)\,, \\
            \tau_4 &= 3\,\Big( (\mathcal{D}\ast\mathcal{D})\times \mathcal{D}_2'(\psi_1) - \frac{\mathcal{S}_2}{\mathcal{S}_1}\,\mathcal{D}\times\mathcal{D}_2'(\psi_1)  \Big)\,.
        \end{split}
        \end{equation}
        where $\mathcal{S}_1$ and $\mathcal{S}_2$ are the fully symmetric Lorentz invariants defined in Eq.~\eqref{two-LI}.
        These $\tau_i$ are fully antisymmetric and have all mass dimension 1.
        In the standard form they can be written as
        \begin{equation}\label{final-S3-basis-1}
        \begin{split}
            \tau_1^{\mu\nu\rho} &=  q_1^\mu\,\delta^{\nu\rho}  + q_2^\nu\,\delta^{\rho\mu} +  q_3^\rho\,\delta^{\mu\nu} \,, \\[1mm]
            \mathcal{S}_0\,\tau_2^{\mu\nu\rho} &=  q_1^\mu\,q_2^\nu\,q_3^\rho \,, \\[2mm]
            \mathcal{S}_0\,\tau_3^{\mu\nu\rho} &=  p_1^2\,q_1^\mu\,\delta^{\nu\rho}  + p_2^2\,q_2^\nu\,\delta^{\rho\mu} +  p_3^2\,q_3^\rho\,\delta^{\mu\nu} \,, \\[1mm]
            -\frac{\mathcal{S}_1}{\mathcal{A}}\,\mathcal{S}_0\,\tau_4^{\mu\nu\rho} &=  \omega_1\,q_1^\mu\,\delta^{\nu\rho}  + \omega_2\,q_2^\nu\,\delta^{\rho\mu} +  \omega_3\,q_3^\rho\,\delta^{\mu\nu} \,,
        \end{split}
        \end{equation}
        where $\omega_i=p_i\cdot q_i=-p_j^2+p_k^2$ and $\mathcal{A}$ is the antisymmetric variable
        \begin{equation}
           \mathcal{A} = (\mathcal{D}\ast\mathcal{D}) \times \mathcal{D}  = 3s^2a - a^3 \,.
        \end{equation}
        This choice produces dressing functions that are sufficiently well-behaved.
        In total we have arrived at four totally antisymmetric basis elements; consequently, their momentum dressing functions must be totally symmetric
        and can only depend on the Lorentz-invariants $\mathcal{S}_0$, $\mathcal{S}_1$ and $\mathcal{S}_2$.

        Applying a full transverse projection to these four structures does not change their symmetry properties
        because the projection operator is Bose-symmetric.
        With the abbreviations
        \begin{equation}
            \mathcal{T}_i^{\mu\nu} = T^{\mu\alpha}_{p_j}\,T^{\alpha\nu}_{p_k}\,, \quad
            t_i^\mu = T^{\mu\nu}_{p_i}\,q_i = T^{\mu\nu}_{p_i}\,(p_j-p_k)^\nu\,,
        \end{equation}
        where $\{i,j,k\}$ is again an even permutation of $\{1,2,3\}$ and $T_k^{\mu\nu}$ the usual transverse projector defined below Eq.~\eqref{ghost-gluon},
        the final transverse basis elements can be written as
        \begin{equation}\label{final-S3-basis}
        \begin{split}
            \tau_{1\perp}^{\mu\nu\rho} &=  t_1^\mu\,\mathcal{T}_1^{\nu\rho}  + t_2^\nu\,\mathcal{T}_2^{\rho\mu} +  t_3^\rho\,\mathcal{T}_3^{\mu\nu} \,, \\[1mm]
            \mathcal{S}_0\,\tau_{2\perp}^{\mu\nu\rho} &=  t_1^\mu\,t_2^\nu\,t_3^\rho \,, \\[2mm]
            \mathcal{S}_0\,\tau_{3\perp}^{\mu\nu\rho} &=  p_1^2\,t_1^\mu\,\mathcal{T}_1^{\nu\rho}  + p_2^2\,t_2^\nu\,\mathcal{T}_2^{\rho\mu} +  p_3^2\,t_3^\rho\,\mathcal{T}_3^{\mu\nu} \,, \\[1mm]
            -\frac{\mathcal{S}_1}{\mathcal{A}}\,\mathcal{S}_0\,\tau_{4\perp}^{\mu\nu\rho} &=  \omega_1\,t_1^\mu\,\mathcal{T}_1^{\nu\rho}  + \omega_2\,t_2^\nu\,\mathcal{T}_2^{\rho\mu} +  \omega_3\,t_3^\rho\,\mathcal{T}_3^{\mu\nu} \,. \\[1mm]
        \end{split}
        \end{equation}
        Hence, the transversely projected three-gluon vertex is given by
        \begin{equation}\label{3g-transverse-total}
           \Gamma_{3g}^{\mu\nu\rho}(p_1,p_2,p_3) = \sum_{i=1}^{4}F_i(\mathcal{S}_0,\mathcal{S}_1,\mathcal{S}_2)\,\tau_{i\perp}^{\mu\nu\rho}(p_1,p_2,p_3)\,.
        \end{equation}

\subsection{Relation with Ball-Chiu basis}

        Ball and Chiu wrote the tensor decomposition of the three-gluon vertex in the following way~\cite{Ball:1980ax} (see also \cite{Kim:1979ep}):
        \begin{equation}
            \Gamma^{\mu\nu\rho}(p_1,p_2,p_3) = \sum_{i=1}^6 \Gamma_i^{\mu\nu\rho}(p_1,p_2,p_3)\,.
        \end{equation}
        The six antisymmetric vertex contributions are given by
        \begin{equation}\label{Ball-Chiu-basis}
        \begin{split}
            \Gamma_1^{\mu\nu\rho} &= A_1\,q_1^\mu\,\delta^{\nu\rho} + A_2\,q_2^\nu\,\delta^{\rho\mu} + A_3\,q_3^\rho\,\delta^{\mu\nu}\,,  \\[1mm]
            \Gamma_2^{\mu\nu\rho} &= B_1\,p_1^\mu\,\delta^{\nu\rho} + B_2\,p_2^\nu\,\delta^{\rho\mu} + B_3\,p_3^\rho\,\delta^{\mu\nu}\,,  \\[1mm]
            \Gamma_3^{\mu\nu\rho} &= C_1\,q_1^\mu\,t^{\nu\rho}_{23} + C_2\,q_2^\nu\,t^{\rho\mu}_{31} + C_3\,q_3^\rho\,t^{\mu\nu}_{12}\,, \\[1mm]
            \Gamma_4^{\mu\nu\rho} &= S\,\big(p_2^\mu\,p_3^\nu\,p_1^\rho + p_3^\mu\,p_1^\nu\,p_2^\rho\big)\,, \\[1mm]
            \Gamma_5^{\mu\nu\rho} &= F_1\,b_1^\mu\,t^{\nu\rho}_{23} + F_2\,b_2^\nu\,t^{\rho\mu}_{31} + F_3\,b_3^\rho\,t^{\mu\nu}_{12}\,,  \\[1mm]
            \Gamma_6^{\mu\nu\rho} &= H\,\big( p_2^\mu\,p_3^\nu\,p_1^\rho - p_3^\mu\,p_1^\nu\,p_2^\rho \\
                                  & \qquad + b_1^\mu\,\delta^{\nu\rho} + b_2^\nu\,\delta^{\rho\mu} + b_3^\rho\,\delta^{\mu\nu}\big)\,.
        \end{split}
        \end{equation}
        As before, $q_i = p_j - p_k$ and we abbreviated
        \begin{equation}\label{analytic-projectors}
        \begin{split}
            t^{\mu\nu}_{ij} &= p_i\cdot p_j\,\delta^{\mu\nu} - p_j^\mu\,p_i^\nu\,, \\
            b_i^\mu &= \tfrac{1}{2}\,t^{\mu\alpha}_{ii}\,q_i^\alpha = p_i\cdot p_j\,p_k^\mu - p_k\cdot p_i\,p_j^\mu\,.
        \end{split}
        \end{equation}
        $\{i,j,k\}$ is again a cyclic permutation of $\{1,2,3\}$, and $p_i + p_j + p_k = 0$.
        The `projector' $t^{\mu\nu}_{ij}$ is transverse to $p_i^\mu$ and $p_j^\nu$, and $b_i^\mu$ is transverse to $p_i^\mu$.
        The elements $\Gamma_5$ and $\Gamma_6$ are transverse in all indices.
        The dressing functions $A_i$, $B_i$, $C_i$, $S$, $F_i$ and $H$ are scalar functions of the arguments $p_1^2$, $p_2^2$ and $p_3^2$, and
        $A_1$, $A_2$, $A_3$ are even permutations of each other.

       \begin{figure*}[!t]
                  \begin{center}
                  \includegraphics[scale=0.37]{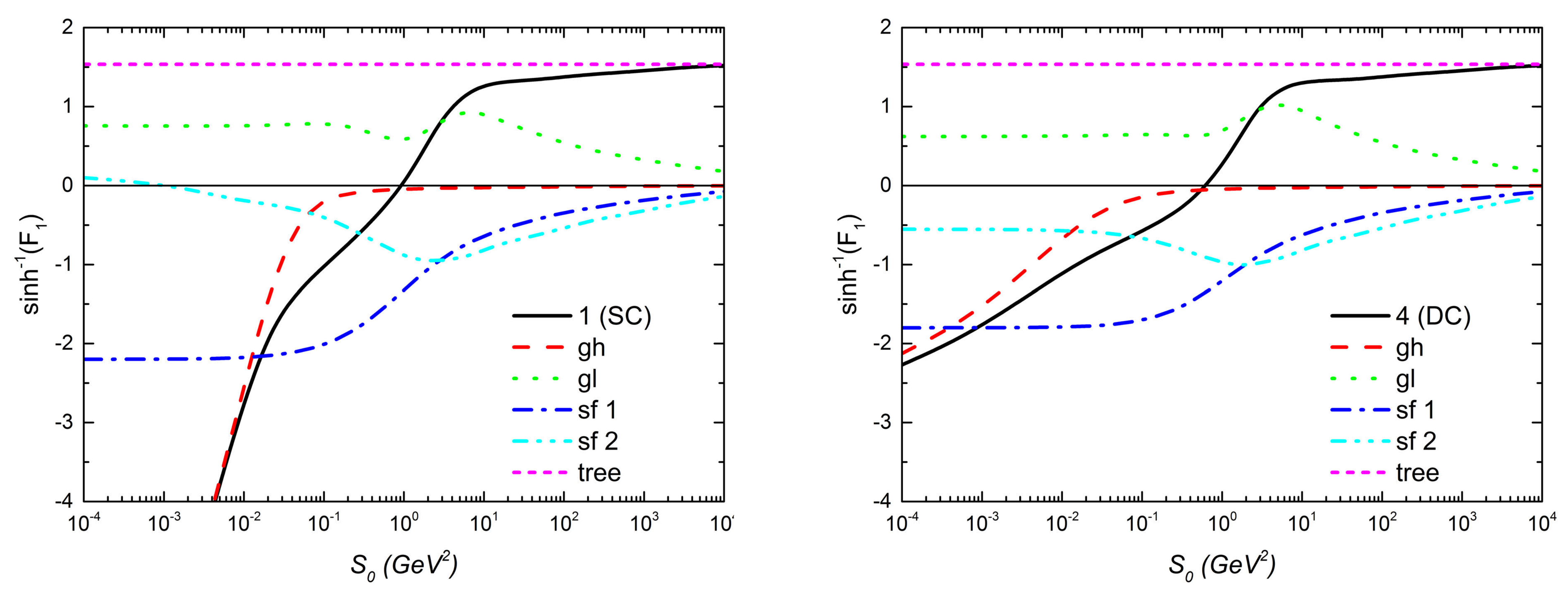}
                  \caption{Individual contributions to the tree-level dressing $F_1$ for scaling (\textit{left panel})
                  and decoupling (\textit{right panel}). The names of the individual contributions follow from Eq.~\eqref{3g-DSE}.}\label{fig:3gv-breakdown}
                  \end{center}
      \end{figure*}

        In order to extract a Bose-antisymmetric
        tensor basis from Eq.~\eqref{Ball-Chiu-basis}, one has to work out the permutation-group multiplets for the dressing functions.
        $A_i$, $C_i$ and $F_i$ are symmetric in their first two momentum arguments, hence they produce symmetric singlets $\mathcal{S}$ and doublets of type $\mathcal{D}_1$.
        The dressing function $B_i$ is antisymmetric in its first two arguments, so it generates an antisymmetric singlet $\mathcal{A}$ and a doublet of type $\mathcal{D}_2$.
        $S$ is fully antisymmetric and $H$ is fully symmetric in all arguments.
        This will produce a tensor decomposition of similar structure as in Table~\ref{tab:permutation-group-basis}.
        For example, if we use $A_3$ as the permutation-group seed, we obtain the following Lorentz-invariant dressing functions via Eq.~\eqref{perm-multiplets}:
        \begin{equation*}
        \begin{split}
            \mathcal{S}(A_3) &= 2\,(A_1+A_2+A_3)\,, \\
            \mathcal{A}(A_3) &= 0\,, \\
            \mathcal{D}_1(A_3) &= \left[ \begin{array}{c} a_1(A_3) \\ s_1(A_3) \end{array}\right] = \left[ \begin{array}{c} 2\,(A_1-A_2) \\ -\tfrac{2}{\sqrt{3}}\,(A_1+A_2-2A_3) \end{array}\right], \\
            \mathcal{D}_2(A_3) &= 0\,.
        \end{split}
        \end{equation*}
        Upon inverting these relations, the
        Ball-Chiu structure $\Gamma_1$ can be written as
        \begin{equation}\label{Gamma-1-BC}
        \begin{split}
            \Gamma_1 &= \tfrac{1}{6}\,\mathcal{S}(A_3)\,\mathcal{A}'(\psi_1) + \tfrac{1}{4}\,\mathcal{D}_1(A_3)\times \mathcal{D}_2'(\psi_1) \,,
        \end{split}
        \end{equation}
        where the tensor structures $\mathcal{A}'(\psi_1)$ and $\mathcal{D}_2'(\psi_1)$ are those in Table~\ref{tab:permutation-group-basis}.
        One can further expand $\mathcal{D}_1(A_3)$ in terms of the two momentum doublets $\mathcal{D}=\binom{a}{s}$ and $\mathcal{D}\ast\mathcal{D}$ according to Eq.~\eqref{perm-decomposition-doublet},
        so that its two dressing functions $a_1(A_3)$ and $s_1(A_3)$ become linear combinations of two fully symmetric singlet functions:
        \begin{equation}
            \mathcal{D}_1(A_3) = \mathcal{S}'(A_3)\,\mathcal{D} + \mathcal{S}''(A_3)\,\mathcal{D}\ast\mathcal{D}\,.
        \end{equation}
        The Ball-Chiu structure $\Gamma_1$ then accommodates three independent
        Bose-symmetric dressing functions $\mathcal{S}(A_3)$, $\mathcal{S}'(A_3)$ and $\mathcal{S}''(A_3)$ with corresponding tensor structures
        \begin{equation}
             \mathcal{A}'(\psi_1)\,, \quad  \mathcal{D}\times \mathcal{D}_2'(\psi_1)\,, \quad (\mathcal{D}\ast\mathcal{D})\times \mathcal{D}_2'(\psi_1)\,,
        \end{equation}
        which are just those in Eq.~\eqref{D,D*D}.
        
        Similarly, the the Ball-Chiu structure $\Gamma_2$ takes the form
        \begin{equation}
             \Gamma_2 = \tfrac{1}{6}\,\mathcal{A}(B_3)\,\mathcal{S}'(\psi_6) - \tfrac{1}{4}\,\mathcal{D}_2(B_3)\times \mathcal{D}_1'(\psi_6)\,, 
        \end{equation}
        and the remaining relations for $\Gamma_3 \dots \Gamma_6$ are collected in Appendix~\ref{sec:bc-basis}.

\section{Results and discussion} \label{sec:results}

    Let us briefly recapitulate our setup.
    We solved the truncated three-gluon DSE, the symmetrized version of Eq.~\eqref{3g-DSE} and Fig.~\ref{fig:dse3gtrunc}.
    We used DSE solutions for the ghost and gluon propagators (Sec.~\ref{sec:propagators}), a bare ghost-gluon vertex,
    and a model for the four-gluon vertex (Sec~\ref{sec:four-gluon-vertex}) as input.
    The scaling and decoupling solutions are continuously connected via the propagator input; for presentation purposes
    we will restrict ourselves to the decoupling set 4 (Table~\ref{tab:GZ-fits}) which is close to the `extreme' decoupling case.
    It is sufficient to retain the transverse projection of the vertex, Eq.~\eqref{3g-transverse-total}, which depends
    on four Lorentz-invariant dressing functions $F_i(\mathcal{S}_0,\mathcal{S}_1,\mathcal{S}_2)$. The momentum variables are discussed in Sec.~\ref{sec:lorentz-invariants}.
    In order to provide instructive figures that can accommodate sign changes and IR divergences,
    we plot $\sinh^{-1} F_i$ instead of $F_i$ (we recall that $\sinh^{-1} x \approx x$ for $x \lesssim 1$ whereas it grows logarithmically for large $|x|$).

\subsection{Symmetric momentum configuration}

    We start with the symmetric momentum configuration $a=s=0$, which corresponds to $p_1^2=p_2^2=p_3^2$.
    Fig.~\ref{fig:3gv-breakdown} shows the various contributions to the dominant tree-level dressing function $F_1$ as a function of the symmetric variable $\mathcal{S}_0$.
    The curves correspond to the diagrams in Fig.~\ref{fig:dse3gtrunc}: tree-level, ghost loop, gluon loop, and the two swordfish diagrams, which add up to the final result given by the solid line.
    The different signs of the individual contributions already indicate that the system is subject to delicate cancellation effects
    which makes it a numerically highly non-trivial problem.

       \begin{figure*}[!t]
                  \begin{center}
                  \includegraphics[scale=0.35]{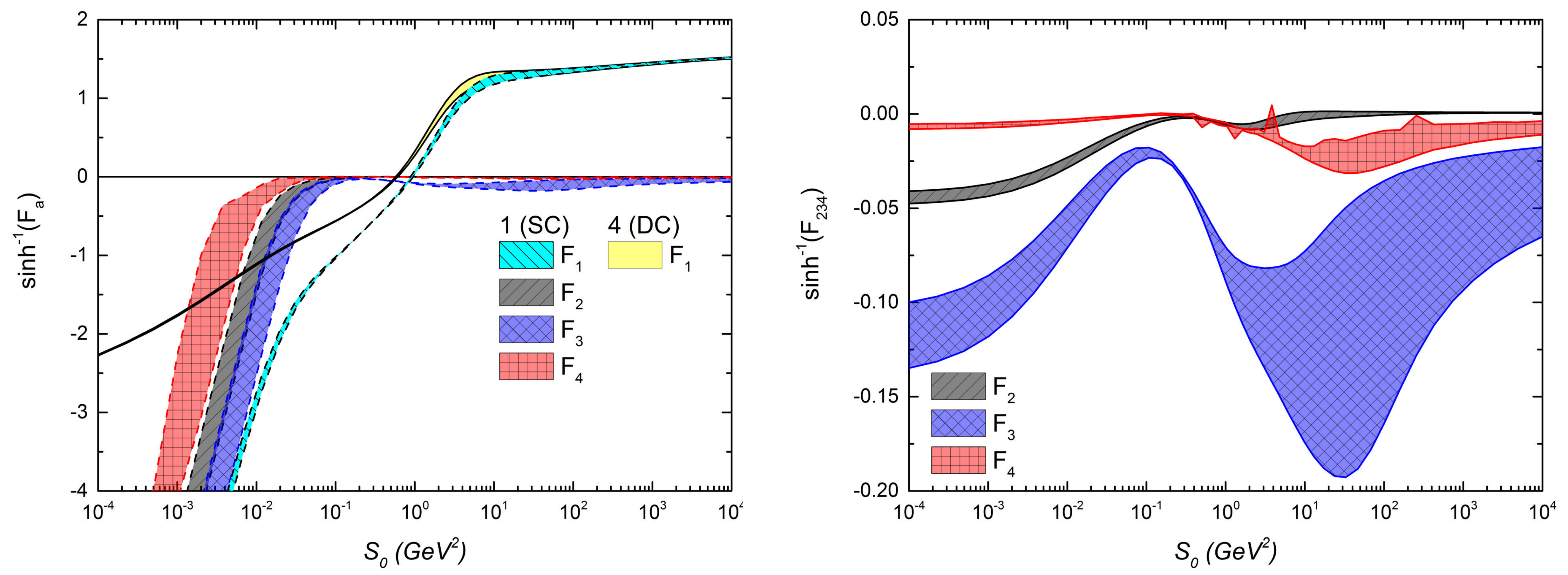}
                  \caption{Angular dependence (indicated by bands) of $F_i$ as a function of $\mathcal{S}_0$. In the left panel
                  we contrast scaling with decoupling, where the former features a stronger angular dependence
                  in the IR. In the right-panel, we show the angular dependence of $F_2$ to $F_4$ for
                  decoupling.}\label{fig:angularSCDC}
                  \end{center}
      \end{figure*}

    For large momenta, $F_1$ approaches the renormalization constant $Z_1=Z_3/\widetilde{Z}_3$ which is fixed from the propagator input\footnote{Instead
    of the values in Table~\ref{tab:GZ-fits} we employ a fixed ghost renormalization constant $\widetilde{Z}_3^\text{SC}=1.529$ in our calculation, which corresponds to the scaling solution.
    This choice is consistent because all the ghost propagators, also the decoupling ones, approach the same value $1/\widetilde{Z}_3^\text{SC}$ at the numerical cut-off.}.
    The diagram that drives the IR (but is otherwise negligible) is the negative ghost-loop contribution.
    It diverges with $-3\kappa$ in the scaling case~\cite{Alkofer:2004it,Huber:2007kc} and logarithmically in the decoupling solution~\cite{Aguilar:2013vaa,Pelaez:2013cpa}.
    Concerning the remaining diagrams, the gluon loop is positive and produces a bump in the mid-momentum region.
    This bump is partially cancelled by the negative swordfish diagrams which overwhelm the gluon loop and shift the total sum further in the negative direction,
    thereby producing a zero crossing in $F_1$. Remarkably, the zero crossing does not happen in the deep IR
    but rather at a hadronic scale $\mathcal{S}_0 \sim 1$~GeV$^2$.

    A sign change for $F_1$ has been anticipated in lattice calculations~\cite{Cucchieri:2008qm} and found in recent continuum studies~\cite{Huber:2012zj,Aguilar:2013vaa,Pelaez:2013cpa,Blum:2014gna}.
    However, all these works predict a zero crossing deep in the IR region. This can happen for various reasons;
    from Fig.~\ref{fig:3gv-breakdown} it is clear that a tree-level plus ghost-loop only calculation will produce a sign change at very low momenta.
    As we will discuss in more detail in Sec.~\ref{sec:rg-improvement}, a certain choice of renormalization-group improved vertices can also have a
    sizeable impact on the location of the zero crossing. We have checked that the choice of projector that has been used in lattice
    studies has no impact on our result. Nevertheless, lattice data on the three-gluon vertex are so far only available for two-color QCD;
    it remains to be seen whether forthcoming $SU(3)$ results will confirm such a behavior or not.

    It is worthwhile to note that both scaling and decoupling solutions are essentially identical except for the deep IR region.
    This may not come as a surprise since already the propagators show the same behavior, but it nurtures the speculation whether the
    distinction between scaling and decoupling has any measurable physical relevance.
    In any case, the zero of $F_1$ at $\mathcal{S}_0 \sim 1$~GeV$^2$ is a robust feature in both scenarios\footnote{Notice, however, the factor $1/6$ in the definition of $\mathcal{S}_0$, Eq.~\eqref{S0-a-s}.}.
    Of course one cannot exclude the possibility that the DSE ingredients which are currently modelled
    (the four-gluon vertex) or discarded (the last row in Fig.~\ref{fig:dse3g})
    can have a quantitative impact on the location of the sign change.

\subsection{Angular dependence}



           \begin{figure*}[!t]
                  \begin{center}
                  \includegraphics[scale=0.33]{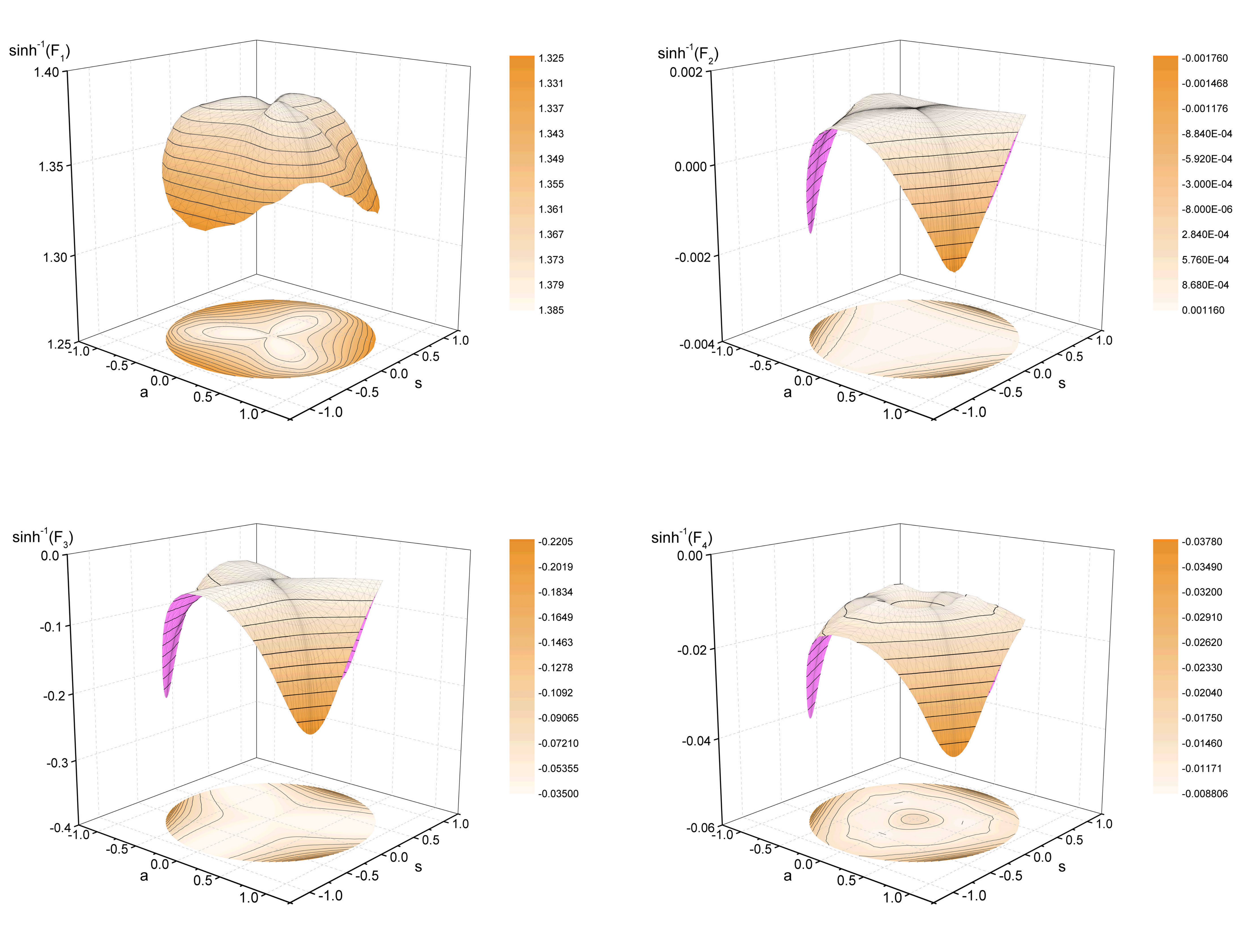} 
                  \caption{Angular dependence of the $F_i$ in the $(a,s)$ plane at fixed $\mathcal{S}_0=102$~GeV$^2$. The soft-divergence enhancements in
                  $F_2$, $F_3$ and $F_4$ can be clearly seen, whereas in $F_1$ the angular dependence is essentially flat.}\label{fig:3gv-soft-sing}
                  \end{center}
      \end{figure*}

    In the left-panel of Fig.~\ref{fig:angularSCDC} we show all four dressing functions for scaling, contrasted with the leading component in decoupling.
    The angular dependence is represented by the shaded regions and is significantly larger in the $F_2$, $F_3$ and $F_4$ components.  
    However, these components are themselves suppressed when compared to the tree-level dressing; the next relevant component is $F_3$ which contributes at the $\sim 10\%$ level.
    These results justify a restriction to the tree-level structure for modelling purposes. It is also evident that, for scaling, all components feature the same scaling behavior in the IR.

    In the right-panel of Fig.~\ref{fig:angularSCDC} we focus upon the sub-leading components in the decoupling scenario.
    The scaling results are practically identical except in the IR $\mathcal{S}_0 \lesssim (0.5\,\text{GeV})^2$ where they begin to deviate and eventually diverge.
    Note the strong angular dependence in the dressing functions, particularly in the UV, whose correct description is essential to obtain numerically stable solutions.
    It is produced by a strong enhancement at the three soft kinematic points discussed earlier in connection with Figs.~\ref{fig:phasespace} and~\ref{fig:phasespace2}.

      This behavior is apparent in Fig.~\ref{fig:3gv-soft-sing}, which shows all four $F_i$ in the $(a,s)$ plane at a fixed value of $\mathcal{S}_0=102$~GeV$^2$.
      We exemplify decoupling set 4, but at this scale the dressing functions are essentially identical for scaling and decoupling solutions.
      As anticipated in Sec.~\ref{sec:lorentz-invariants}, the dressing functions recover the rotationally symmetric profiles illustrated in Fig.~\ref{fig:phasespace2}.
      Close to the symmetric point, the sub-leading dressing functions
      are suppressed compared to $F_1$, whereas for larger radii their magnitudes change.
      While $F_1$ shows almost no angular dependence, the other functions exhibit a
      soft gluon enhancement in the three corners where $p_i^2=0 \Leftrightarrow \mathcal{S}_2=-1$.
      The qualitative behavior is similar if $\mathcal{S}_0$ is taken to be in the mid-momentum region, although the peaks become sharper in the scaling case.
      Whether the subleading dressing functions diverge or not is numerically difficult to resolve; soft singularities are expected to happen at least in the scaling case. 

      We found that these features can also have an impact on the stability of the DSE iteration.
      Without accounting for the Bose symmetry of the phase space, which is manifest in the $(a,s)$ plane,
      soft singularities can show up in seemingly random places and complicate the numerical solution process.
      To resolve this we defined our grid directly in the variables $\mathcal{S}_0$, $a$ and $s$.
      It is convenient to express $a$ and $s$ in cylindrical coordinates and perform a Chebyshev expansion in the angular variable.
      A fully symmetric dressing function is thus symmetric within any slice of $120^\circ$,
      and the pattern is repeated by going around the circle once. Hence, it is sufficient to calculate only one third of the $(a,s)$ plane.

    \subsection{Running coupling from three-gluon vertex}

        \renewcommand{\arraystretch}{1.2}

\begin{figure}[!t]
                  \begin{center}
                  \includegraphics[scale=0.36]{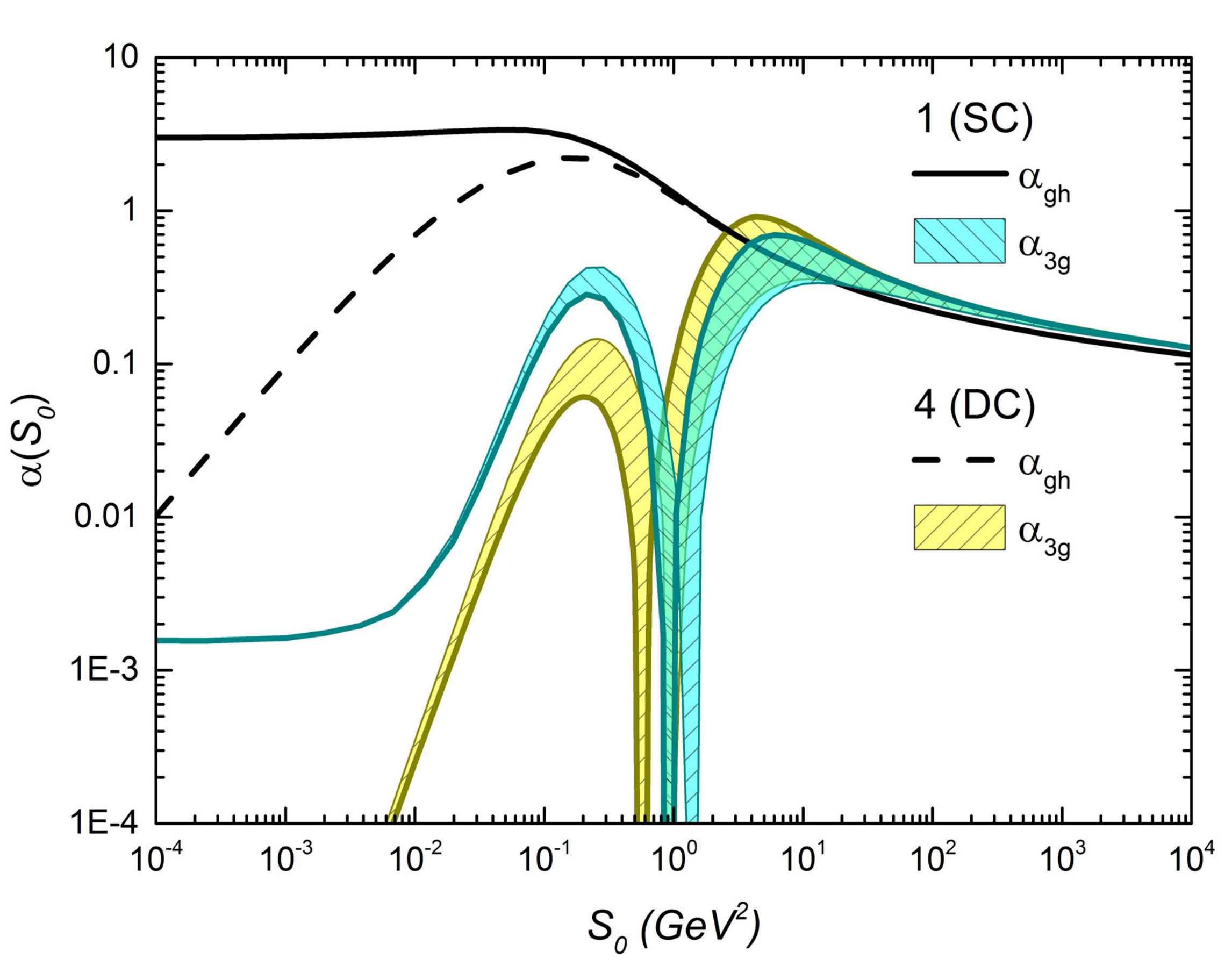}
                  \caption{The running coupling of the three-gluon vertex, Eq.~\eqref{eqn:running3g} with $n=2$, for both scaling- and decoupling-type scenarios, together with that of the ghost-gluon vertex. }\label{fig:3gv-coupling}
                  \end{center}
      \end{figure}

      Fig.~\ref{fig:3gv-coupling} shows the running couplings from the ghost-gluon and three-gluon vertices, defined as
      \begin{equation}
      \begin{array}{rl}
          \alpha_\text{gh}(p^2) &= \alpha(\mu^2)\,Z(p^2)\,G^2(p^2)\,,  \\
          \alpha_\text{3g}(p^2) &= \alpha(\mu^2)\,Z^3(p^2)\,F_1^2(p^2)\,,
      \end{array}
      \quad \alpha(\mu^2) = \frac{g^2(\mu^2)}{4\pi}\,.
      \end{equation}
      They are renormalization-group invariant and scale with an inverse logarithm in the UV,
      as can be inferred from Eqs.~(\ref{GFs-dressed-vs-bare}--\ref{Z-relations-2}) and Table~\ref{tab:GFs-IR-UV}.
      From the definition of $\alpha_\text{3g}(p^2)$ it is clear that the zero crossing in $F_1$ will be inherited by the running coupling,
      which is positive due to its quadratic dependence on $F_1$.
      Both couplings agree in the UV but their non-perturbative shape is quite different, which underlines the fact
      that there is no unique `non-perturbative running coupling'.

      In fact, one can construct many renormalization-group invariant combinations of $Z$, $G$ and higher $n-$point functions
      which satisfy the same properties:
         \begin{align}
         \alpha_\text{gh} &=\frac{g^2}{4\pi}\, Z\,(G\,\Gamma_\text{gh})^2\,, \\
         \alpha^{(n)}_\text{3g} &=\frac{g^2}{4\pi}\, Z\,(G\,\Gamma_\text{gh})^2\left[\frac{Z\,\Gamma_\text{3g}}{G\,\Gamma_\text{gh}}\right]^n, \label{eqn:running3g}\\
         \alpha^{(n)}_\text{4g} &=\frac{g^2}{4\pi}\, Z\,(G\,\Gamma_\text{gh})^2\left[\frac{Z\,\Gamma_\text{4g}}{(G\,\Gamma_\text{gh})^2}\right]^n.
         \end{align}
      $\Gamma_\text{gh}$, $\Gamma_\text{3g}$ and $\Gamma_\text{4g}$ denote the tree-level dressing functions of the ghost-gluon, three-gluon and four-gluon vertex, respectively.
      From Eqs.~(\ref{GFs-dressed-vs-bare}--\ref{Z-relations-1}), their renormalization-group invariance also holds for $\widetilde{Z}_1\neq 1$,
      and they all have the same UV scaling with an inverse logarithm.
         Hence, they are all equally valid definitions of `non-perturbative running couplings', although their
         shape in the small-momentum region will be very different.
         In the IR they all become constant in the scaling case; we find $\alpha^{(2)}_\text{3g}\simeq0.0016$ (see Fig.~\ref{fig:3gv-coupling}).
         For decoupling they vanish as $\alpha_\text{gh} \sim p^2$ or $\alpha_\text{3g}$, $\alpha_\text{4g}\sim (p^2)^{n+1}$.
         Usually $\alpha_\text{gh}$, $\alpha_\text{3g}^{(2)}$ and $\alpha_\text{4g}^{(1)}$ are quoted; for the latter two, these choices of $n$ eliminate the dependence on the ghost-gluon vertex dressing.

\subsection{Model and truncation dependence} \label{sec:rg-improvement}

       \begin{figure}[!t]
                  \begin{center}
                  \includegraphics[scale=0.36]{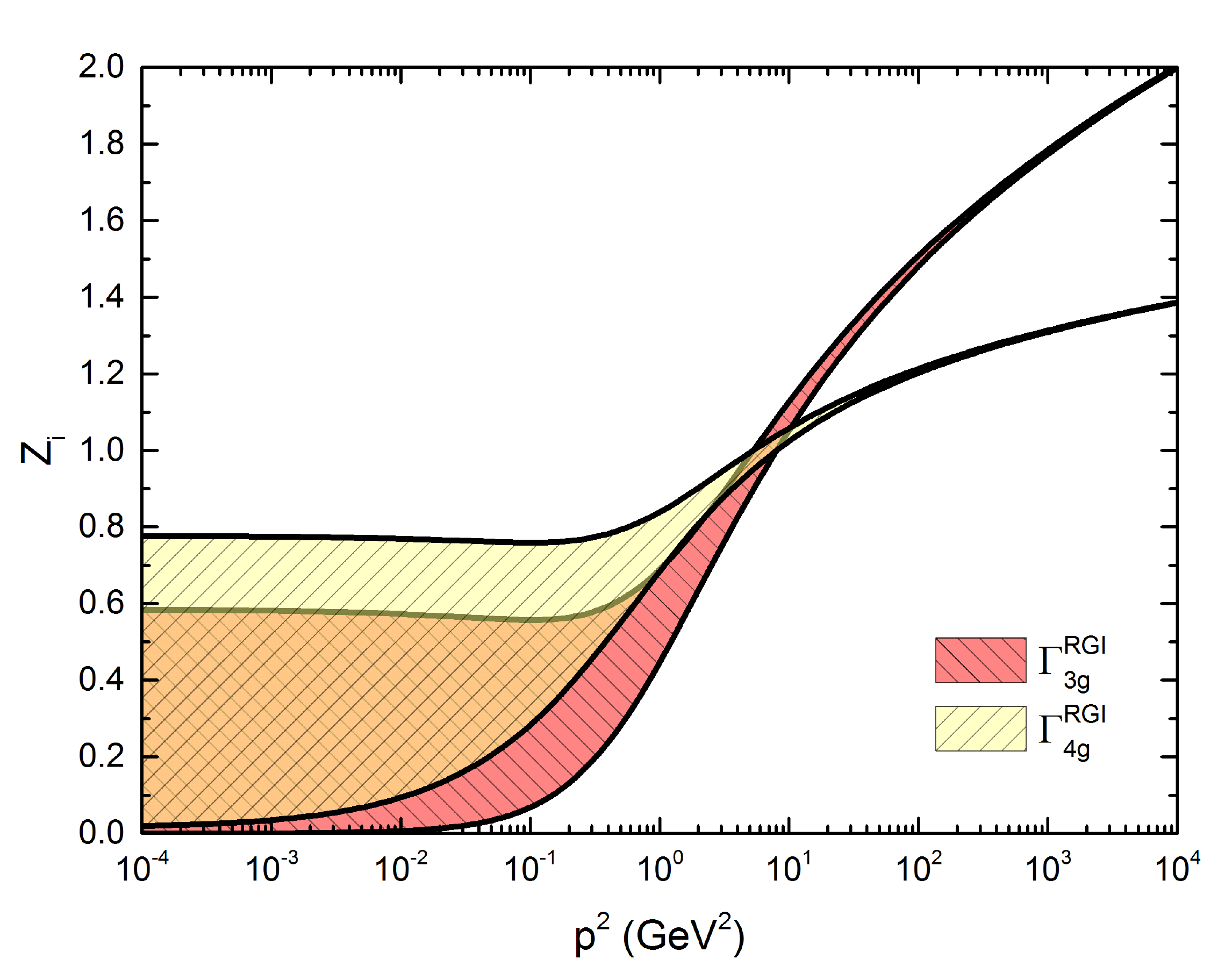}
                  \caption{The spread in the RG improved vertices for scaling- vs. decoupling-type solutions. }\label{fig:3g4gspread}
                  \end{center}
      \end{figure}

      It is non-trivial to reproduce the correct anomalous dimensions in the UV at the level of one-loop diagrams within the DSE framework, since in each diagram one vertex is always bare. It is then the higher order diagrams (in particular those at two-loop) that provide this consistency. A commonly used technique is to effectively dress each bare vertex with a `renormalization group (RG) improvement'. The idea is to construct combinations of the ghost and gluon propagator dressings, $G$ and $Z$, such that the correct anomalous dimensions of the vertex are reproduced together with being a finite constant in the IR. That is, following Ref~\cite{Huber:2012kd,Blum:2014gna} one could introduce a momentum dependence in the renormalization constants $Z_1$ and $Z_4$:
       \begin{figure*}[!t]
                  \begin{center}
                  \includegraphics[scale=0.37]{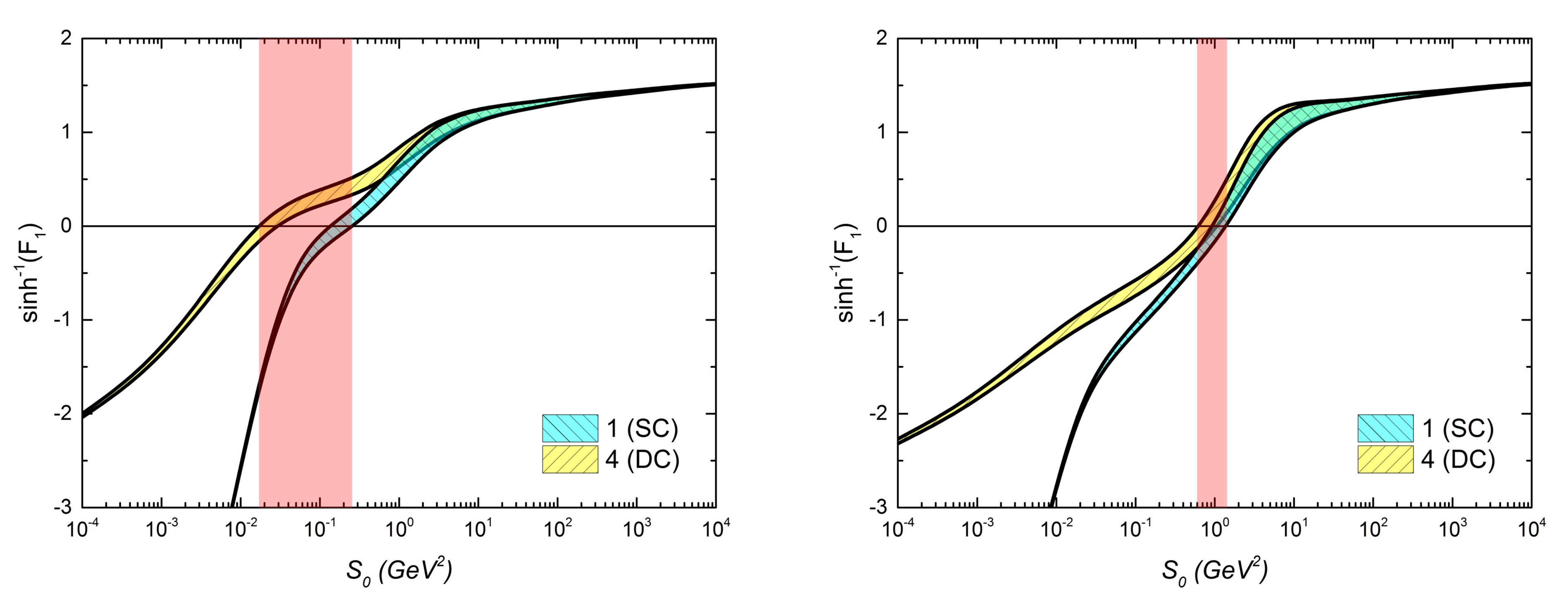}
                  \caption{Tree-level dressing function $F_1$ with (\textit{left panel}) and without (\textit{right panel}) RG improvement. The bands indicate the model dependence of the four-gluon vertex. The two vertically-shaded regions mark the spread in the location of the zero crossing between scaling and decoupling.}\label{fig:3gv-RG}
                  \end{center}
      \end{figure*}
      \begin{align}\label{eqn:rgimprovement}
      	Z_1\rightarrow Z^{a_1}G^{a_2}\;,\qquad      	Z_4\rightarrow Z^{b_1}G^{b_2}\;,
      \end{align}
      where $a_1=0$, $a_2=-17/9$, $b_1=0$, $b_2=-8/9$ for decoupling, and for scaling $a_1=-1+3\delta$, $a_2=2a_1$, $b_1=-1+4\delta$, $b_2=2b_1$ with $\delta=-9/44$.

      In Fig.~\ref{fig:3g4gspread} we show the functional form of this RG improvement for various forms of propagator input. At large perturbative momentum, as expected, the deviation is small. However, it modifies the IR and mid-momentum regions far too strongly. Since these RG improved vertices
      feature in the gluon-loop and swordfish diagrams only, they provide a suppression therein with respect to the ghost-loop diagram; essentially they provide a momentum dependent re-weighting of the contributions, eventually leading back to ghost-loop dominance.  Moreover, since these dressings apply to only one external leg they explicitly break bose-symmetry; in our system, however, this is mitigated through symmetrization of the DSE.

      In Fig.~\ref{fig:3gv-RG} we show a comparison of the leading $F_1$ component for scaling and decoupling, both with and without the RG improvement. The IR and far UV are essentially the same since one region is dominated by perturbative effects, whilst the other is determined by ghost-dominance which is independent of the RG dressings. However, we see a very strong dependence in the mid-momentum (owing to the effective momentum dependent reweighting) which changes the location of the zero crossing by between one and two orders of magnitude in $\mathcal{S}_0$. Without RG improvement, the location of the zero crossing is very similar for both scaling and decoupling. However, with the RG improvement an order of magnitude shift is introduced; this suggests that the use of
      \eqref{eqn:rgimprovement} introduces an additional model uncertainty.

      Another type of model dependence comes from the four-gluon vertex.
      The bands in Fig.~\ref{fig:3gv-RG} represent the parametric dependence of our four-gluon vertex dressing discussed in connection with Eq.~\eqref{4g-dressing} and shown in Fig.~\ref{fig:4gv-dressing}.
      We chose our model so that the deviation between scaling and decoupling starts to set in at a momentum scale $\approx 10$~GeV$^2$, which is relatively high compared to the propagators where the difference appears
      only much further down in the IR. Therefore, the spread in $F_1$ between scaling and decoupling in the vicinity of the zero crossing is essentially due to the difference in the four-gluon vertex model.
      This makes clear that the location of the zero crossing will certainly depend on the truncation, i.e., the (full) four-gluon vertex and neglected diagrams in the DSE.
      Still, both model versions in Fig.~\ref{fig:4gv-dressing} lead to a sign change at $\mathcal{S}_0 \sim 1$~GeV$^2$ which is a robust feature within our present truncation.


%

We note that a dressed four-gluon vertex (instead of a bare one) also turned out to be essential for the stability of the DSE solution.
We found during the course of these investigations that taking the four-gluon vertex to be bare throughout leads
to non-convergence of the DSE for the three-gluon vertex. The non-linear integral equations were solved using
standard iteration in combination with under-relaxation; to confirm the non-convergence we further employed
Newton's iterative method. Neither techniques led to stable solutions. We also searched for solutions in which the
three-gluon vertex features no zero crossing, without success.
This indicates that the dressing of the four-gluon vertex is important, and moreover that it must be sufficiently
strong in order to provide the needed stability into the system.
Our vertex model in Fig.~\ref{fig:4gv-dressing} provides that strength.
At this stage, it is hard to judge what impact
the missing diagrams have on the system; certainly they play a role in further stabilising the equations, but may have
a material impact on the mid-momentum region and hence the location of the zero-crossing.

Our propagator input is consistent with solutions of the ghost and gluon propagator in which two-loop terms are neglected. This has the effect that the bump in the gluon propagator does not have the same strength as seen on the lattice. To test whether this has a material impact on our truncation, we enhanced the gluon propagator by multiplication with the function
\begin{align}
1+x\exp(-x)\;,
\end{align}
where $x=p^2/\Lambda^2$ and $\Lambda$ is the same scale as obtained for the propagator fits. This increases the peak in the running coupling of the ghost-gluon vertex by $1/3$, but has no discernible impact on the location of the zero crossing of the three-gluon vertex. This leads us to believe that it is indeed the details of the four-gluon vertex and missing contributions that are of import.

We also explored the 3PI system~\cite{Berges:2005hc} in which all vertices are dressed, but the four-gluon vertex is always bare and no two-loop terms appear.
We found that this system was unstable without a small enhancement of $Z_4$. Qualitatively, however, the solutions are similar to the 1PI system with only small modifications in the IR.

\section{Conclusions and outlook} \label{sec:conclusions}

We provided the first calculation of the three-gluon vertex within the framework of the Dyson--Schwinger equations in which the full covariant structure of the vertex is back-coupled. They scale as expected in the uniform IR limit (logarithmically or with a power) in both decoupling and scaling. In the sub-leading components we found enhancements due to soft-collinear divergences.

The presence of a zero crossing in the leading component was confirmed. The DSE solution shifts its value from the deep infrared
to a scale $\sim 1$ GeV; however, its location is dependent upon the modelling of the four-gluon vertex and missing components.
The presence of a zero crossing may have a tangible impact on hadron physics, in particular bound-state studies, dependent on whether it occurs at a high enough scale to be relevant. Here, excited states would be more sensitive; this can be exemplified through Bethe--Salpeter studies beyond rainbow ladder featuring a dressed quark-gluon vertex~\cite{Fischer:2009jm}. We anticipate that future lattice calculations for
$SU(3)$ may provide an answer.

It would be interesting to incorporate unquenching effects into this system, since they will obviously have a qualitative impact. Along these lines, large-$N_f$ calculations and applications to Technicolor within the Dyson--Schwinger framework can be explored.

\section*{Acknowledgements}

         We are grateful to Christian S. Fischer, Markus Q. Huber, Mario Mitter and Lorenz von Smekal for valuable discussions and a critical reading of the manuscript.
We acknowledge support by the German Science Foundation (DFG) under project number DFG TR-16, the Austrian Science Fund (FWF) under project numbers M1333-N16, J3039-N16 and P25121-N27, and from the Doktoratskolleg ``Hadrons in Vacuum, Nuclei and Stars'' (FWF) DK W1203-N16.

\appendix

\section{Orthonormal tensor basis}  \label{sec:orthonormal-tensor-basis}

            \renewcommand{\arraystretch}{0.5}

             In Section~\ref{sec:bose-symm-tensor-basis} we constructed a tensor basis for the three-gluon vertex
             that implements the features of transversality and Bose symmetry.
             However, for the numerical solution of the three-gluon vertex DSE it is advantageous to work
             with an orthonormal tensor basis (even if it is not necessarily Bose-symmetric) since this reduces the numerical effort considerably.
             We will detail its construction in the following.

             The three-gluon vertex has 14 basis elements in total.
             The simplest construction principle for an orthonormal basis
             starts with the momenta $k$ and $Q$ defined in Eq.~\eqref{momenta-relative-total}, or equivalently
             $k_\pm = k \pm Q/2$, so that
             \begin{equation}
                 k_- = -p_1, \qquad  k_+ = p_2, \qquad Q = -p_3\,.
             \end{equation}
             We can orthonormalize $k$ and $Q$ by defining
             \begin{equation}\label{momenta-sd}
                 d^\mu = \widehat{Q}^\mu\,, \qquad s^\mu = \widehat{k_T}^\mu\,,
             \end{equation}
             where $k^\mu_T= T_Q^{\mu\nu} k^\nu$ is the transverse projection of $k$, with $T_Q^{\mu\nu} = \delta^{\mu\nu} -  \widehat{Q}^\mu\,\widehat{Q}^\nu$,
             and a hat denotes a normalized four-momentum.
             In the frame where
             \begin{align}
                 Q = \sqrt{Q^2}\left(\begin{array}{c} 0 \\ 0 \\ 0 \\ 1 \end{array}\right), \quad
                 k=\sqrt{k^2}\left(\begin{array}{c} 0 \\ 0 \\ \sqrt{1-z^2} \\ z \end{array}\right),
             \end{align}
             $d$ and $s$ are then simply the unit vectors in the $4-$ and $3-$directions, respectively.

             If we temporarily define
             \begin{equation}\label{basis-prelim-3}
             \begin{array}{rl}
                \mathsf{T}_1^{\mu\nu} &= \delta^{\mu\nu}\,, \\
                \mathsf{T}_2^{\mu\nu} &= s^\mu s^\nu\,, \\
                \mathsf{T}_3^{\mu\nu} &= d^\mu d^\nu\,,
             \end{array}\qquad
             \begin{array}{rl}
                \mathsf{T}_4^{\mu\nu} &= s^\mu d^\nu + d^\mu s^\nu \,,\\
                \mathsf{T}_5^{\mu\nu} &= s^\mu d^\nu - d^\mu s^\nu\,,
             \end{array}
             \end{equation}
             we can write down a complete 14-dimensional basis
             by collecting all possible combinations of $s$, $d$ and the Kronecker delta:
             \begin{align}
                  \big\{ s^\rho, \; & d^\rho \big\}  \times \left\{ \mathsf{T}_1^{\mu\nu}, \, \mathsf{T}_2^{\mu\nu}, \, \mathsf{T}_3^{\mu\nu}, \, \mathsf{T}_4^{\mu\nu}, \, \mathsf{T}_5^{\mu\nu}\right\}, \label{basis-prelim-1}\\
                 &\left\{ s^\mu, \,  d^\mu \right\} \times \delta^{\rho\nu}\,, \quad
                 \left\{ s^\nu, \, d^\nu \right\} \times \delta^{\rho\mu}\,. \label{basis-prelim-2}
             \end{align}

             The next step is to construct a basis in terms of $s$ and $d$ with definite transversality properties.
             Since in Landau gauge any internal or external gluon leg of the vertex will always be contracted with a transverse gluon propagator,
             it is sufficient to work with those basis elements that are transverse to all momenta $k_-^\mu$, $k_+^\nu$ and $Q^\rho$.
             If we introduce the auxiliary variables
             \begin{equation}\label{auxiliary-variables}
                 a = \sqrt{3\xi}\,z\,, \qquad b = \sqrt{3\xi}\,\sqrt{1-z^2}\,,
             \end{equation}
             we can write the momenta as
             \begin{equation}\label{Qf}
                 k_\pm^\mu = \sqrt{t}\left( b\,s^\mu + (a \pm 1)\,d^\mu\right),  \quad
                 Q^\mu = 2\sqrt{t}\,d^\mu\,.
             \end{equation}
             The elements with $d^\rho$ in Eq.~\eqref{basis-prelim-1} are longitudinal with respect to $Q^\rho$, whereas those with $s^\rho$ are transverse.
             We can re-express the five elements in~\eqref{basis-prelim-3} in terms of tensor structures
             which have also definite transversality properties with respect to $k_-^\mu$ and $k_+^\nu$.
             These have been worked out in Ref.~\cite{Eichmann:2012mp} in the context of nucleon Compton scattering and they
             read:\footnote{In Ref.~\cite{Eichmann:2012mp}, $\mathsf{Y}_1$ $\dots$ $\mathsf{Y}_5$ correspond to $\mathsf{Y}_1$,
             $\mathsf{Y}_3$, $\mathsf{Y}_{10}$, $\mathsf{Y}_{11}$ and $\mathsf{Y}_{12}$ in Eqs.~(60), (D11) and below (D12).}
             \begin{equation*}\label{qcv-transverse-basis-Y}
             \begin{split}
                \mathsf{Y}_1 &= \frac{1}{\sqrt{2}}\left( \mathsf{T}_1 - \mathsf{T}_2 - \mathsf{T}_3\right), \\
                \mathsf{Y}_2 &= \frac{1}{\sqrt{n_1 n_2}}\left[ (1-a^2)\, \mathsf{T}_2 - b^2\,\mathsf{T}_3 + ab\,\mathsf{T}_4 - b\,\mathsf{T}_5  \right], \\
                \mathsf{Y}_3 &= \frac{1}{\sqrt{n_1 n_2}}\left[ (1-a^2)\, \mathsf{T}_3 - b^2\,\mathsf{T}_2 - ab\,\mathsf{T}_4 - b\,\mathsf{T}_5  \right], \\
                \mathsf{Y}_4 &= \frac{1}{\sqrt{2 n_1 n_2}}\left[ (1-a^2+b^2)\, \mathsf{T}_4 - 2ab\,(\mathsf{T}_2-\mathsf{T}_3)  \right], \\
                \mathsf{Y}_5 &= \frac{1}{\sqrt{2 n_1 n_2}}\left[ (1-a^2-b^2)\, \mathsf{T}_5 +2b\,(\mathsf{T}_2 + \mathsf{T}_3)  \right].
             \end{split}
             \end{equation*}
             Here we omitted the Lorentz indices and abbreviated
             \begin{equation}
                 n_1 = 1+a^2+b^2\,, \qquad n_2 = n_1 - \frac{4a^2}{n_1}\,.
             \end{equation}
             $\mathsf{Y}_1$ and $\mathsf{Y}_2$ are completely transverse in the indices $\mu$ and $\nu$; $\mathsf{Y}_3$ is completely longitudinal, and the remaining ones are mixed.
             Thus, from Eq.~\eqref{basis-prelim-1} we get only two fully transverse elements: $s^\rho\,\mathsf{Y}_1^{\mu\nu}$ and $s^\rho\,\mathsf{Y}_2^{\mu\nu}$.

             In order to make the transversality properties of the
             remaining elements in Eq.~\eqref{basis-prelim-2} manifest, it is helpful to rewrite
             $s^\mu$ and $d^\mu$ in terms of $k_\pm^\mu$ and the momenta $s_\pm^\mu := T_{k_\pm}^{\mu\alpha}\,s^\alpha$ which are transverse to $k_\pm^\mu$.
             If we also normalize them, we arrive at
             \begin{equation}\label{sk-mom}
             \begin{split}
                 \widehat{s_\pm}^\mu &= \frac{1}{\sqrt{n_1 \pm 2a}}\left[ (a \pm 1)\,s^\mu - b\,d^\mu \right], \\
                 \widehat{k_\pm}^\mu &= \frac{1}{\sqrt{n_1 \pm 2a}}\left[ b\,s^\mu +(a \pm 1)\,d^\mu \right].
             \end{split}
             \end{equation}
             If we further replace the Kronecker deltas in Eq.~\eqref{basis-prelim-2} by $\delta^{\rho\nu} \rightarrow \mathsf{Y}_1^{\rho\nu}$ and
             $\delta^{\rho\mu} \rightarrow \mathsf{Y}_1^{\rho\mu}$ (which are transverse to both $s$ and $d$) we arrive at the following
             complete basis:
             \begin{equation}\label{basis-simple}     \renewcommand{\arraystretch}{1.2}
             \begin{split}
             \begin{array}{r@{\!\;}l}
                 \rho_1^{\mu\nu\rho} &= \mathsf{Y}_1^{\mu\nu}\,s^\rho  \\
                 \rho_2^{\mu\nu\rho} &= \mathsf{Y}_2^{\mu\nu}\,s^\rho  \\
                 \rho_3^{\mu\nu\rho} &= \mathsf{Y}_1^{\rho\nu}\,\widehat{s_-}^\mu  \\
                 \rho_4^{\mu\nu\rho} &= \mathsf{Y}_1^{\rho\mu}\,\widehat{s_+}^\nu \\[4mm]
                 \rho_9^{\mu\nu\rho} &= \mathsf{Y}_3^{\mu\nu}\,s^\rho  \\
                 \rho_{10}^{\mu\nu\rho} &= \mathsf{Y}_4^{\mu\nu}\,s^\rho  \\
                 \rho_{11}^{\mu\nu\rho} &= \mathsf{Y}_5^{\mu\nu}\,s^\rho
             \end{array}\qquad
             \begin{array}{r@{\!\;}l}
                 \rho_5^{\mu\nu\rho} &= \mathsf{Y}_1^{\mu\nu}\,d^\rho  \\
                 \rho_6^{\mu\nu\rho} &= \mathsf{Y}_2^{\mu\nu}\,d^\rho  \\
                 \rho_7^{\mu\nu\rho} &= \mathsf{Y}_1^{\rho\nu}\,\widehat{k_-}^\mu  \\
                 \rho_8^{\mu\nu\rho} &= \mathsf{Y}_1^{\rho\mu}\,\widehat{k_+}^\nu \\[4mm]
                 \rho_{12}^{\mu\nu\rho} &= \mathsf{Y}_3^{\mu\nu}\,d^\rho  \\
                 \rho_{13}^{\mu\nu\rho} &= \mathsf{Y}_4^{\mu\nu}\,d^\rho  \\
                 \rho_{14}^{\mu\nu\rho} &= \mathsf{Y}_5^{\mu\nu}\,d^\rho
             \end{array}
             \end{split}
             \end{equation}
             It is already orthonormal because the basis elements satisfy the orthogonality relation
             \begin{equation}
                \rho_i^{\mu\nu\rho}\,\rho_j^{\mu\nu\rho} = \delta_{ij}\,.
             \end{equation}
             Only the first four elements are fully transverse: applying three transverse projectors leaves them invariant
             while eliminating all the remaining ones:
             \begin{align}
                T_{k_-}^{\mu\alpha} \,T_{k_+}^{\nu\beta} \,T_Q^{\rho\gamma} \,\rho_i^{\alpha\beta\gamma} &= \rho_i^{\mu\nu\rho} &&  \qquad i\leq 4\,, \label{transverse-projection}\\
                T_{k_-}^{\mu\alpha} \,T_{k_+}^{\nu\beta} \,T_Q^{\rho\gamma} \,\rho_i^{\alpha\beta\gamma} &= 0 &&  \qquad i >4\,.
             \end{align}
             Therefore it is sufficient to work with these first four alone since they carry the complete dynamics.
             The decoupling of the three-gluon vertex DSE into transverse and longitudinal equations is manifest with this basis choice.
             The $\rho_i^{\mu\nu\rho}$ do not have definite Bose symmetry and neither do their dressing functions.
             However, this is irrelevant for the numerical solution of the DSE as long as the full $(a,s)$ plane from Fig.~\ref{fig:phasespace2} is back-coupled during the iteration.
             The dressing functions $F_i$ attached to the $\tau_{i\perp}$ of Eq.~\eqref{final-S3-basis} are then obtained from those of the $\rho_j$ above through rotation.

\section{Relation with Ball-Chiu basis}  \label{sec:bc-basis}

        In this appendix we return to the relation between the Ball-Chiu basis and Table~\ref{tab:permutation-group-basis}.
        The result for $\Gamma_1$ has been given in Eq.~\eqref{Gamma-1-BC}; here we also
        collect the remaining Ball-Chiu structures.
        To shorten the notation, we abbreviate the tensor basis multiplets of Table~\ref{tab:permutation-group-basis} by
        \begin{equation*}
           \mathcal{A}_1,\, \mathcal{D}_1,\quad \mathcal{A}_2,\quad \mathcal{S}_3,\,\mathcal{D}_3,\quad \mathcal{A}_4,\, \mathcal{D}_4,\quad \mathcal{S}_5,\quad \mathcal{S}_6,\,\mathcal{D}_6
        \end{equation*}
        and those for the Ball-Chiu dressing functions constructed from Eq.~\eqref{Ball-Chiu-basis} by
        \begin{equation*}
           \mathcal{S}_A,\, \mathcal{D}_A,\quad  \mathcal{A}_B,\, \mathcal{D}_B,\quad \mathcal{S}_C,\, \mathcal{D}_C, \quad \mathcal{S}_F,\, \mathcal{D}_F\,,
        \end{equation*}
        whereas $S$ is already antisymmetric and $H$ is symmetric.
        In the following, $\mathcal{S}_0$ and $\mathcal{D}=\binom{a}{s}$ are the usual momentum multiplets,
        where we use the additional shorthand $\widetilde{\mathcal{S}} = \mathcal{D}\cdot\mathcal{D}-1$ and $\widetilde{\mathcal{D}} = \mathcal{D}+\mathcal{D}\ast\mathcal{D}$. Then we obtain:

        \begin{equation}
        \begin{split}
            \Gamma_1 &= \tfrac{1}{6}\,\mathcal{S}_A\,\mathcal{A}_1 + \tfrac{1}{4}\,\mathcal{D}_A\times \mathcal{D}_1 \,, \\[2mm]
            \Gamma_2 &= \tfrac{1}{6}\,\mathcal{A}_B\,\mathcal{S}_6 - \tfrac{1}{4}\,\mathcal{D}_B\times \mathcal{D}_6\,, \\[2mm]
            \Gamma_3 &= \tfrac{1}{8}\,\mathcal{S}_C\,\big(\mathcal{A}_2-\tfrac{1}{3}\,\mathcal{A}_4\big)
                      - \tfrac{1}{16}\,\mathcal{D}_C\times(\mathcal{D}_4-\sqrt{3}\,\mathcal{D}_3) \\
                     &- \tfrac{1}{6}\,\mathcal{S}_0\,(\mathcal{S}_C+\sqrt{3}\,\mathcal{D}\cdot\mathcal{D}_C)\,\mathcal{A}_1 \\
                     &- \tfrac{1}{4}\,\mathcal{S}_0 \,\big( \mathcal{D}_C - \mathcal{D}\ast\mathcal{D}_C + \tfrac{2}{\sqrt{3}}\,\mathcal{D}\,\mathcal{S}_C\big)\times\mathcal{D}_1\,,  \\[2mm]
            \Gamma_4 &= -\tfrac{1}{4} \,S\,\big( \mathcal{S}_3+\mathcal{S}_5\big)   ,
        \end{split}
        \end{equation}
        \begin{equation}
        \begin{split}
            \Gamma_5 &= \tfrac{1}{8}\,\mathcal{S}_0\,\Big[ \big( \mathcal{S}_F-\tfrac{\sqrt{3}}{2}\,\mathcal{D}\cdot\mathcal{D}_F\big)\big( \mathcal{A}_2-\tfrac{1}{3}\,\mathcal{A}_4\big) \\
                     &+ \tfrac{1}{2}\,\mathcal{D}\times\mathcal{D}_F\,(3\mathcal{S}_5-\mathcal{S}_3) \\
                     &- \big( \mathcal{D}\,\mathcal{S}_F - \tfrac{\sqrt{3}}{2}\,\mathcal{D}_F\big)\times\mathcal{D}_3 \\
                     &- \tfrac{1}{\sqrt{3}}\,\big( \mathcal{D}\,\mathcal{S}_F + \tfrac{\sqrt{3}}{2}\,\mathcal{D}_F + \sqrt{3}\,\mathcal{D}\ast\mathcal{D}_F\big) \times \mathcal{D}_4 \Big]\\
                     &+ \tfrac{1}{4}\,\mathcal{S}_0^2\,\Big[ \tfrac{2}{3}\,\big( \widetilde{\mathcal{S}}\,\mathcal{S}_F-\tfrac{\sqrt{3}}{2}\,\widetilde{\mathcal{D}}\cdot\mathcal{D}_F\big)\,\mathcal{A}_1 \\
                     &+ \big( \widetilde{\mathcal{S}}\,\mathcal{D}_F + \tfrac{1}{2}\,\widetilde{\mathcal{D}}\ast\mathcal{D}_F - \tfrac{1}{\sqrt{3}}\,\widetilde{D}\,\mathcal{S}_F\big)\times \mathcal{D}_1\\
                     &+ (\widetilde{\mathcal{D}}\times\mathcal{D}_F)\,\mathcal{S}_6
                      + \big( \widetilde{\mathcal{D}}\,\mathcal{S}_F + \tfrac{\sqrt{3}}{2}\,\widetilde{\mathcal{D}}\ast\mathcal{D}_F\big)\times\mathcal{D}_6   \Big]\,,\\[2mm]
            \Gamma_6 &= H\,\Big[ \mathcal{S}_0 \, \mathcal{A}_1 + \tfrac{1}{4}\,\big( \mathcal{A}_2 + \mathcal{A}_4\big)  \\
                     &-\tfrac{\sqrt{3}}{2}\,\mathcal{S}_0\,\mathcal{D}\times\big( \mathcal{D}_1 + \sqrt{3}\,\mathcal{D}_6\big)  \Big]\,.
        \end{split}
        \end{equation}
        We recall that only the tensor structures $\mathcal{A}_1$, $\mathcal{D}_1$ and $\mathcal{A}_2$ will survive a full transverse projection with three gluon propagators.
        This entails that the Ball-Chiu structures $\Gamma_2$ and $\Gamma_4$ vanish upon such a projection;
        the four dressing functions they contain do not carry any physics (in Landau gauge).
        The six functions in $\Gamma_1$ and $\Gamma_3$ are constrained by the STI, whereas the four functions in $\Gamma_5$ and $\Gamma_6$
        are fully transverse and subject to analyticity constraints. These are enforced by the projectors in Eq.~\eqref{analytic-projectors} which are free of kinematic singularities.
        After a transverse projection, all 10 independent functions collapse into the four structures in Eq.~\eqref{final-S3-basis}.

      \bibliographystyle{apsrev4-1-mod}
      \bibliography{lit}

\end{document}